\documentclass[journal]{IEEEtran}
\usepackage{amsmath,amsfonts,amssymb}
\usepackage{algorithm}
\usepackage{algorithmic}

\usepackage{array}
\usepackage{bm}
\usepackage{subfigure} %
\usepackage{textcomp}
\usepackage{stfloats}
\usepackage{url}
\usepackage{verbatim}
\usepackage{graphicx}
\usepackage{cite}
\makeatletter

\renewcommand{\maketag@@@}[1]{\hbox{\m@th\normalsize\normalfont#1}}%

\makeatother

\allowdisplaybreaks[2]
\newcommand{\figwidth}{8}
\newcounter{MYtempeqncnt}
\begin{document}

\title{6D Movable Antenna Enhanced Multi-Access Point Coordination via Position and Orientation Optimization}

\author{Xiangyu Pi,
	Lipeng Zhu,~\IEEEmembership{Member,~IEEE,}
	Haobin Mao,
	Zhenyu Xiao,~\IEEEmembership{Senior Member,~IEEE,}\\
	Xiang-Gen Xia,~\IEEEmembership{Fellow,~IEEE}
	and Rui Zhang,~\IEEEmembership{Fellow,~IEEE}
\thanks{ X.~Pi, H. Mao, and Z.~Xiao  are with the School of Electronic and Information Engineering, and the State Key Laboratory of CNS/ATM,
Beihang University, Beijing 100191, China. (e-mail: pixiangyu@buaa.edu.cn, maohaobin@buaa.deu.cn, xiaozy@buaa.edu.cn). \emph{(Corresponding author: Zhenyu Xiao)}}
\thanks{L.~Zhu is with the Department of Electrical and Computer Engineering, National University of Singapore, Singapore 117583, Singapore. (e-mail: zhulp@nus.edu.sg).}
\thanks{X.-G. Xia is with the Department of Electrical and Computer Engineering, University of Delaware, Newark, DE 19716, USA. (e-mail: xxia@ee.udel.edu).}
\thanks{R. Zhang is with School of Science and Engineering, Shenzhen Research Institute of Big Data, The Chinese University of
	Hong Kong, Shenzhen, Guangdong 518172, China (e-mail:rzhang@cuhk.edu.cn). He is also with the Department of Electrical
	and Computer Engineering, National University of Singapore, Singapore 117583 (e-mail: elezhang@nus.edu.sg).}

\vspace{-2em}
}

\maketitle
\thispagestyle{empty}
\begin{abstract}
The effective utilization of unlicensed spectrum is regarded as an important direction to enable the massive access and broad coverage for next-generation wireless
local area network (WLAN). Due to the crowded spectrum occupancy and dense user terminals (UTs), the conventional fixed antenna (FA)-based access points (APs) face huge challenges in realizing massive access and interference cancellation.
To address this issue,  in this paper we develop  a six-dimensional movable antenna (6DMA) enhanced multi-AP coordination system for coverage enhancement and interference mitigation.
First, we model the wireless channels between the APs and UTs to characterize their variation with respect to 6DMA movement, in terms of both the three-dimensional (3D) position and 3D orientation of each distributed AP's antenna. Then, an optimization problem is formulated to maximize the weighted sum rate of multiple UTs for their uplink transmissions by jointly optimizing the antenna position vector (APV), the antenna orientation matrix (AOM), and the receive combining matrix over all coordinated APs, subject to the constraints on local antenna movement regions. 
To solve this challenging non-convex optimization problem, we first transform it into a more tractable Lagrangian dual problem. Then, an alternating optimization (AO)-based algorithm is developed by iteratively optimizing the APV and AOM, which are designed by applying the successive convex approximation (SCA) technique and Riemannian manifold optimization-based algorithm, respectively. Moreover, 
to further reduce the overhead of antenna movement, we propose an offline solution for APV and AOM design based on statistical channel state information (CSI). In addition, we further extend the proposed scheme from uni-polarized to dual-polarized modes for all antennas. 
Simulation results show that the proposed 6DMA-enhanced multi-AP coordination system can significantly enhance network capacity, and both of the online and offline 6DMA schemes can attain considerable performance improvement compared to the conventional FA-based schemes.
\end{abstract}


\section{Introduction}
\IEEEPARstart{W}{ith} the explosive growth in data traffic and user terminal (UT) density, the inadequate licensed spectrum is becoming increasingly crowded
to meet the demand for high capacity and massive access in future wireless communication networks \cite{Unlicensed2021,Unlicensed2023}. 
It is forecast that the mobile data traffic will increase with a compound annual growth rate of about $20\%$ through 2029 \cite{traffic2024}, and there are around $70\%$ of the global internet traffic crossing wireless
local area network (WLAN).
To improve network performance, the exploration in the unlicensed spectrum has been recognized to play a crucial role in supporting massive communication and ubiquitous connectivity\cite{WiFi2018,WiFi2024}. 
However, realizing massive access in the unlicensed spectrum practically encounters significant challenges due to the limited spectrum resources and severe interference collisions. Therefore, it is important to explore advanced wireless technologies in the unlicensed spectrum to provide low latency and high reliability for emerging applications such as augmented reality (AR), virtual reality (VR), and industrial Internet of Things (IoT) \cite{AR2016,AR2022}.

Since these applications over WLAN are far beyond the exiting systems such as Wi-Fi 6, a new amendment standard IEEE 802.11be for Wi-Fi 7 has emerged with advanced medium access control (MAC) layer and physical layer (PHY) techniques~\cite{WiFi7Survey,WiFi7Access}, including multiple resource units (multi-RU) support, 4096 quadrature amplitude modulation (4096-QAM), expanded bandwidth of more than 160 MHz, multiple link operations, etc.
One of the key feature differentiating Wi-Fi 7 from Wi-Fi 6 is referred to as multiple access point (multi-AP) coordination (e.g.,
multi-AP joint transmission/reception)~\cite{multiAP1,multiAP2}. Wi-Fi 6 only supports transmission to/from a single AP and spatial reuse between APs and UTs without coordination,
which significantly limits the efficiency of utilizing time, frequency and spatial resources. In addition, in the unlicensed frequency band, the incumbent UTs generally employ the carrier sense multiple access with collision avoidance (CSMA/CA) protocol to access channels in a competitive manner~\cite{CSMA}. This protocol manages network traffic by allowing devices to sense if the channel is occupied and then avoid collisions in sending their data. However, such a competitive mechanism in CSMA/CA protocol inevitably results in impeding channel access fairness and degrading network efficiency. To cope with these challenges,  Wi-Fi 7 exploits the cooperation of multiple APs to support sharing data and control information among APs via wired or wireless links~\cite{SurveymultiAP}, thus improving the spectrum efficiency, enhancing the network coverage and mitigating the interference more effectively over Wi-Fi 6.

Considering that conventional fixed antennas (FAs) are widely implemented in the WLAN, their limited spatial degrees of freedom (DoFs) in antenna deployment may hinder the full exploitation of spatial multiplexing and beamforming gains. On one hand, due to the random mobility of UTs, they may be located at specific positions experiencing poor channel conditions with the FA-based APs. This not only diminishes the signal strength but also hampers the overall spectrum efficiency of the network. On the other hand, high channel correlation between different UTs leads to a degradation in spatial multiplexing performance. The multiple access capacity may be  compromised due to limited spatial separation achieved with FAs. This correlation limits the ability to distinguish between the signals from/to different UTs, thereby increasing multiuer interference and reducing the total data rate.


To overcome the fundamental limitations of conventional FAs, movable antenna (MA), also known as fluid antenna system~\cite{history2024}, has recently been recognized as a novel solution to exploit additional DoFs in antenna movement for improving wireless channel conditions and thus achieving better spatial diversity, beamforming, and/or multiplexing performance~\cite{zhu2023movable, MagazineMA,FA1,FA2,FA2024}. In addition, there have been preliminary works validating that local movements of MAs can yield considerable performance improvements for both narrowband
and wideband communication systems~\cite{zhu2022modeling, zhu2024wideband}. The MA-aided point-to-point multiple-input multiple-output (MIMO) communication systems were studied in~\cite{ma2022mimo} and~\cite{chen2023joint} based on the instantaneous channel state information (CSI) and  statistical CSI, respectively, where the optimization of MAs' positions at transceivers can efficiently increase the MIMO communication capacity. The MA-enabled multiuser-MIMO (MU-MIMO) systems were also investigated in~\cite{MUMA2023ZL,Conf2023Pi}. Therein, the implementation of MAs at the user side and/or base station (BS) side can effectively improve multiple access channel (MAC) conditions and suppress multuser interference. Besides, the MA-aided satellite communication system was proposed in~\cite{zhu2024satellite}, where the reconfiguration of MA array geometry can significantly decrease interference leakage and enhance satellite beam coverage. Since the aforementioned optimization of MAs' positions relies on the acquisition of CSI, the channel estimation for MA systems was investigated in~\cite{ma2023compressed,Cao2023compressed} to reconstruct the channel's multipath components using a finite number of channel measurements. With additional spatial DoFs through antenna rotation, a new six-dimensional MA (6DMA) architecture was introduced in~\cite{6DMA20204model,6DMA20204opt,6DMAhistory}. Therein, the 6DMAs' positions and rotations at the BS are optimized within continuous regions or selected from possible discrete sets to maximize the MU-MIMO network capacity.
6DMA system channel estimation and 6DMA-enhanced wireless sensing were also studied in~\cite{6DMAestimation} and~\cite{6DMAsensing}, respectively. To reduce the implementation cost of 6DMA, a new BS architecture composed of both FAs and MAs was proposed in~\cite{6DMAFA}.  
Besides, the application of MA technology has been further extended to physical-layer security (PLS)~\cite{PLSMA}, intelligent reflecting surface (IRS)~\cite{IRSMA}, unmanned aerial vehicle (UAV) communications~\cite{UAVMA}, integrated communication and sensing (ISAC)~\cite{ISACMA}, over-the-air computation~\cite{OverairMA}, etc. However, all the aforementioned works focus on centralized MAs for wireless systems.

Motivated by the highest DoFs in antenna movement for 6DMAs~\cite{6DMA20204model,6DMA20204opt,6DMAhistory}, in this paper we propose a 6DMA-enhanced multi-AP coordination system for enhancing communication coverage and more effectively mitigating multiuser interference in Wi-Fi networks. Due to the fixed positions and orientations, FAs cannot fully unleash the potential of multi-AP coordination in wireless coverage and multiuser access. In comparison, the utilization of 6DMA in multi-AP coordination can effectively improve the multiuser channel condition and further expand the communication coverage. Moreover, the flexible movement of antennas can significantly decrease the channel correlation and thus enable the access of more UTs in the limited frequency bandwidth.
The main contributions of this paper are summarized as follows: 
\begin{enumerate}
	\item We propose a novel 6DMA-enhanced multi-AP coordination system, in which each AP is equipped with a 6DMA to improve the rate performance of multiple UTs. The wireless channels between APs and UTs are characterized as the functions of the antenna position vector (APV) and the antenna orientation matrix (AOM). Then, we formulate an optimization problem to maximize the weighted sum rate (WSR) of uplink UTs by jointly optimizing the APV, AOM of the distributed 6DMAs at APs and their receive combining matrix, subject to the constraints on local antenna movement regions.
	
	\item To solve the formulated non-convex optimization problem with highly coupled variables, we first transform it into a more tractable Lagrangian dual problem by introducing auxiliary variables to approximate the original problem. Then, an alternating optimization (AO)-based algorithm is developed
	for iteratively optimizing the APV and AOM. Specifically, with the available  instantaneous CSI, the optimal receive combining is derived with the minimum mean-square-error (MMSE) receiver. The locally optimal solutions for APV and AOM are obtained by applying the successive convex
	approximation (SCA) technique and Riemannian manifold optimization-based algorithm, respectively. 
	
	\item To reduce the overhead of antenna movement, we develop an offline solution for APV and AOM design based on statistical CSI. As such, the offline-designed positions and orientations of 6DMAs will remain constant until the statistical channel characteristics change. In addition, we further extend the proposed 6DMA-enhanced multi-AP coordination system from uni-polarized to dual-polarized modes for all antennas, where the corresponding optimization problem can still be efficiently solved by the proposed solution for APV and AOM optimization.

	\item Extensive simulations are presented to evaluate the communication performance of our proposed 6DMA-enhanced multi-AP coordination system. Compared to the conventional FA-based multi-AP coordination, the proposed schemes can significantly suppress multiuser interference and improve the spatial multiplexing performance. Moreover, it is validated that the offline solution for APV and AOM design based on statistical CSI can obtain a considerable performance improvement over FAs with both uni-polarized and dual-polarized antennas. 
\end{enumerate} 

The rest of this paper is organized as follows. In Section~\ref{sec_SystemModel}, we introduce the 6DMA-enhanced multi-AP coordination system model and formulate the 
optimization problem. In Section~\ref{sec_ProposedSolution}, we present the proposed solution for the
formulated optimization problem and discuss its convergence and computational complexities. 
Section~\ref{sec_SimulationResults} shows the simulation results. Finally, the paper is concluded in Section~\ref{sec_Conclusion}.
\begin{figure}[!t]
	\centering
	\includegraphics[width=\figwidth cm]{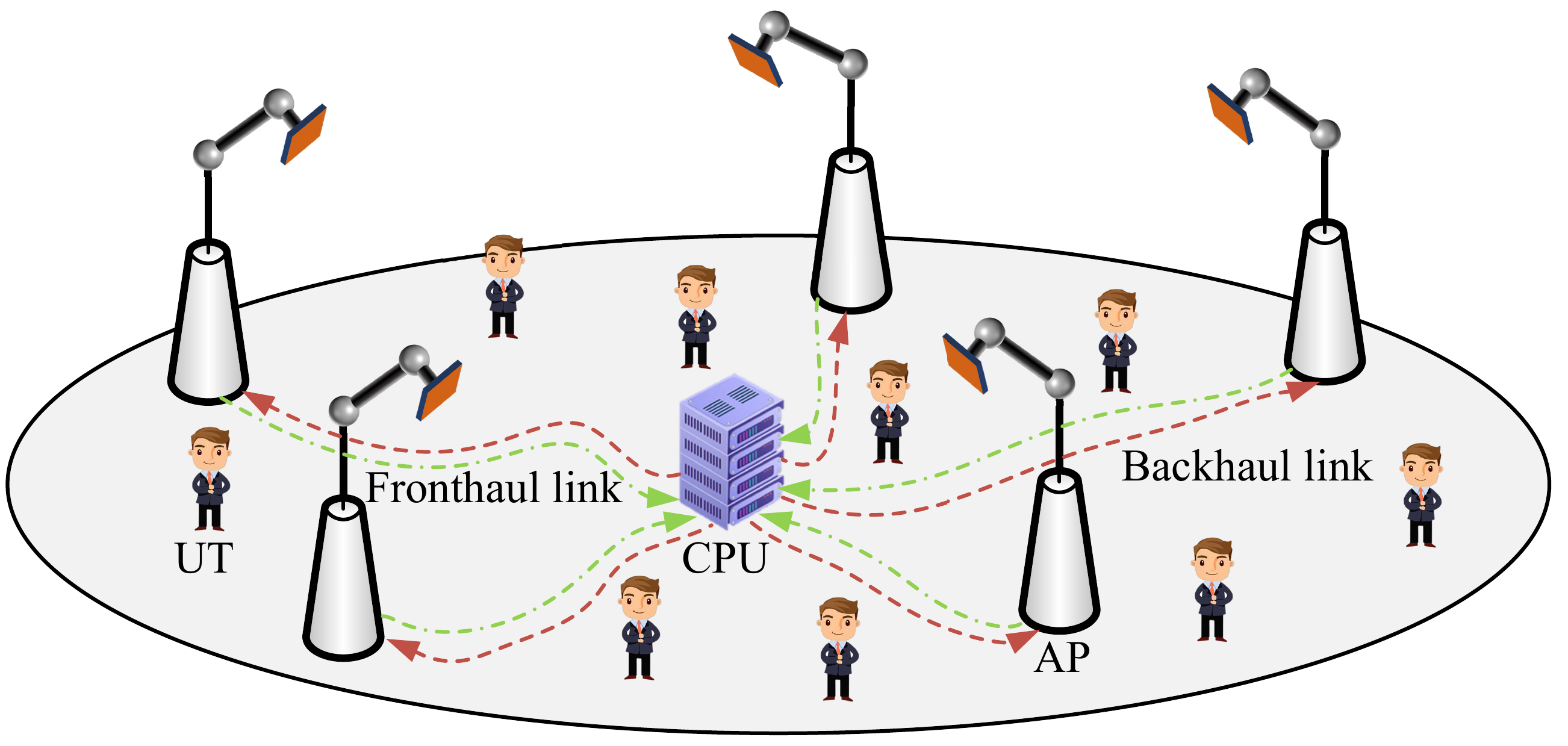}
	\caption{Illustration of the proposed 6DMA-enhanced multi-AP coordination system.}
	\label{fig:system_model}
\end{figure}

\textit{Notation}: $a$, $\mathbf{a}$, $\mathbf{A}$, and $\mathcal{A}$ denote a scalar, a vector, a matrix, and a set, respectively. $\mathbb R$ and $\mathbb C$ represent the sets of real and complex numbers, respectively. 
$(\cdot)^{\rm{T}}$, $(\cdot)^{*}$, $(\cdot)^{\rm{H}}$, and $(\cdot)^{\dagger}$ denote transpose, conjugate, conjugate transpose, and Moore-Penrose inverse, respectively. 
$\mathcal{A}\backslash\mathcal{B}$ represents the set subtraction of
$\mathcal{B}$ from $\mathcal{A}$. $|\mathcal{A}|$ denotes the cardinality of set $\mathcal{A}$. 
$[\mathbf{a}]_i$ and $[\mathbf{A}]_{i,j}$ denote the $i$-th entry of vector $\mathbf{a}$ and the entry in the $i$-th row and $j$-th column of
matrix $\mathbf{A}$, respectively.
$|\cdot|$, $\angle(\cdot)$, and $\Re\{\cdot\}$ denote the amplitude, the phase and the real part of a complex
number, respectively.
$||\mathbf a||_2$ denotes the $l_2$-norm of vector $\bf a$. $||\mathbf A||_2$ and $||\mathbf A||_F$ denote the $l_2$-norm and Frobenius norm of matrix $\mathbf A$, respectively.
$\mathbf A \succeq 0$ represents that $\mathbf A$ is a positive semi-definite matrix.
$\partial(\cdot)$ denotes the partial differential of a function. $\nabla f(\bf x)$ and $\nabla^2 f(\bf x)$ represent the
gradient vector and hessian matrix of function $f(\bf x)$, respectively.
$\mathbf{I}_N$ and $\mathbf 0_{N}$ denote the identity matrix and null matrix of size $N\times N$, respectively. $\mathcal{CN}(0,\mathbf{A})$ represents the circularly symmetric
complex Gaussian (CSCG) distribution with mean zero and covariance matrix $\mathbf{A}$. 
$\mathcal{U}[a,b]$ denotes the uniform distribution over the real-number interval $[a,b]$.

\section{System Model and Problem Formulation}\label{sec_SystemModel}
As shown in Fig.~\ref{fig:system_model}, in this paper we consider a 6DMA-enhanced multi-AP coordination communication system, where $M$ single-6DMA APs cooperatively serve $K$ single-FA UTs. 
For ease of notation, the sets of APs and UTs are defined as $\mathcal{M}$ and $\mathcal K$, respectively. 
Different from the omnidirectional antennas typically assumed in existing works, the directional antennas are installed at APs with directivity gain. For the time being, we assume the antennas at the UTs and APs are uni-polarized, while the case of employing dual-polarized antennas at the APs will be considered later (cf. Section \ref{Sec:dual}). Note that all APs are connected to the same central processing unit (CPU), which allows for payload data to be exchanged between the CPU and APs via fronthaul links, and controls each 6DMA's position and orientation through backhaul links. Consequently, the CPU enables all APs to communicate with all UTs simultaneously over the same time-frequency resources through spatial multiplexing.
In addition, each 6DMA can be flexibly moved with the aid of motors to adjust both its three-dimensional (3D) position and its 3D orientation~\cite{6DMA20204model}.

\subsection{Antenna Movement}
\begin{figure}[!t]
	\centering
	\includegraphics[scale = 0.21]{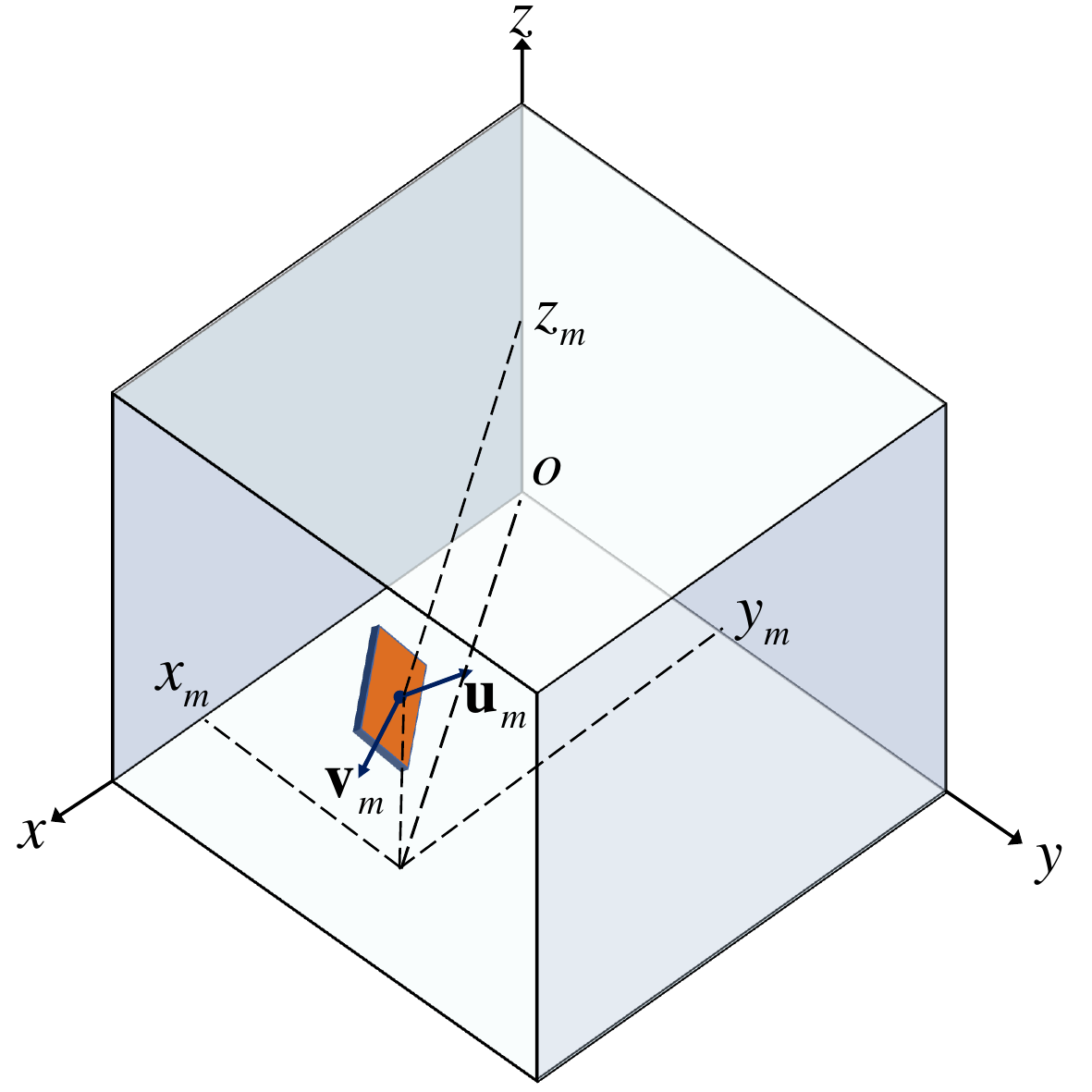}
	\caption{Representation of antenna's position and orientation in the LCCS.}
	\label{fig:LCCS}
\end{figure}

 For ease of exposition, we develop a 3D local Cartesian coordinate system (LCCS) at the $m$-th AP shown in Fig.~\ref{fig:LCCS}, with one vertex of the moving region being the reference point. Consequently, the 6DMA's position is given by
\begin{equation}
	\label{Antenna_position}
	\mathbf{q}_m=[x_m,y_m,z_m]^\mathrm{T}\in\mathbb R^{3}.
\end{equation}
Without loss of generality, the antenna moving regions of all APs are assumed to be  cuboids, i.e., $\mathcal C_m = [x^{\text{min}},x^{\text{max}}]\times[y^{\text{min}},y^{\text{max}}]\times [z^{\text{min}},z^{\text{max}}]$.
To describe the 6DMA's orientation, the two-way vector representation is introduced, which can be expressed as
\begin{equation}
	\label{Antenna_orientation}
	\mathbf{A}_m=[\mathbf{u}_m,\mathbf{v}_m]\in\mathbb R^{3\times 2},
\end{equation}
where $\mathbf{u}_m\in\mathbb R^{3}$ is the normal vector of 6DMA plane shown in Fig.~\ref{fig:LCCS} and $\mathbf{v}_m\in\mathbb R^{3}$ is a normalized vector in the 6DMA plane. From this definition, $\mathbf{u}_m$ and $\mathbf{v}_m$ are a pair of mutually orthogonal unit vectors, i.e., $\|\mathbf{u}_m\|_2=\|\mathbf{v}_m\|_2=1$ and $\mathbf{u}_m^\text{T}\mathbf{v}_m = 0$, which is equivalent to  $\mathbf{A}_m^\text{T}\mathbf{A}_m = \mathbf{I}_2$. 
Then, all 6DMAs' positions and orientations can be represented by APV, i.e, ${\tilde{\mathbf q}} = {\left[ {{\bf{q}}_1^{\rm{T}},{\bf{q}}_2^{\rm{T}}, \ldots ,{\bf{q}}_M^{\rm{T}}} \right]^{\rm{T}}}$ and AOM, i.e., ${\tilde{\mathbf A}} = [{{\bf{A}}_1},{{\bf{A}}_2}, \ldots ,{{\bf{A}}_M}]$, respectively.
Thus, there are in total six DoFs for each 6DMA's movement in the 3D space, and the effects of 6DMAs' positions and orientations on wireless channels are modeled in the next.

\subsection{Channel Model}
For 6DMA-enhanced multi-AP coordination systems, the channel response is determined by the positions and orientations of the antennas as well as the propagation environments.
Since the effect of antenna polarization was neglected in existing 6DMA channel models~\cite{6DMA20204model,6DMA20204opt}, we provide in this paper a more accurate 6DMA channel model by considering the antenna polarization and extending the field-response channel model for MAs in~\cite{zhu2022modeling,ma2022mimo}. Considering the multi-path propagation environment from each UT to each 6DMA, the corresponding channel response is given by the superposition of multiple channel paths. In addition, the far-field condition between each AP and each UT is assumed to be guaranteed, since the signal propagation distance is much larger than the size of antenna movement region at each AP. Thus, for each channel path component, its angle of arrival (AoA) is fixed for different antenna positions/orientations. The phase of the complex path coefficient varies with different antenna positions, since the signal propagation distance is only determined by the antenna positions under the same AoA. On the other hand, the amplitude of the complex path coefficient is affected by the antenna orientation, since the antenna gain is not uniform in all directions and it varies depending on the antenna orientation.

For the given antenna orientation ${{\bf{A}}_m}$, we first characterize the impact of antenna position on the channel response. For the incident wave from UT $k$ to AP $m$ at the $l$-th path, with elevation AoA $\theta_{k,m}^l$ and azimuth AoA $\phi_{k,m}^l$, its wave vector in the LCCS is given by
\begin{equation}
	\small
	\mathbf{d}(\mathbf{\psi}_{k,m}^l)=[\cos\theta_{k,m}^l\cos\phi_{k,m}^l,\cos\theta_{k,m}^l\sin\phi_{k,m}^l,\sin\theta_{k,m}^l]^{\mathrm{T}},
\end{equation}
where $\mathbf{\psi}_{k,m}^l = (\theta_{k,m}^l,\phi_{k,m}^l)$ denotes the wave direction. Thus, the phase difference of the $l$-th receive path between $\mathbf q_m$ and reference point at LCCS is $2\pi\mathbf{d}(\mathbf{\psi}_{k,m}^l)^{\mathrm{T}}\mathbf{q}_m/\lambda$,
where $\lambda$ is the carrier wavelength. Denote $L_{k,m}$ as the number of  total channel paths from UT $k$ to AP $m$. Then, we define the field-response vector (FRV) of the $m$-th 6DMA for receiving a signal from UT $k$ as~\cite{zhu2022modeling}
\begin{equation}\label{FRV}
	{\bf{f}}_{k,m}({\bf{q}}_m) = {\left[ {{e^{j\frac{{2\pi }}{\lambda }{\bf{d}}{{\left( {{\bf{\psi }}_{k,m}^l} \right)}^{\rm{T}}}{{\bf{q}}_m}}}} \right]_{1\le l \le L_{k,m}}^{\rm{T}}}.
\end{equation}
As a result, the channel response between UT $k$ and AP $m$ is given by
\begin{equation}\label{channel}
	h_{k,m}({{\bf{q}}_m},{{\bf{A}}_m}) = {\bf{f}}_{k,m}^{\rm{H}}({{\bf{q}}_m}){{\bf{G}}_{k,m}}({{\bf{A}}_m}){{\bf{a}}_{k,m}},
\end{equation}
where ${{\bf{a}}_{k,m}} = {[ {a_{k,m}^1,a_{k,m}^2, \ldots ,a_{k,m}^{{L_{k,m}}}} ]^{\rm{T}}}$ is path-response vector (PRV), which represents the multi-path response coefficients from UT $k$ to the reference point of AP $m$ with unit antenna gain, and is determined by the propagation environment. 
${{\bf{G}}_{k,m}}({{\bf{A}}_m})$ is a diagonal matrix with each diagonal element representing the effective antenna gain of the corresponding path. Noteworthy, ${{\bf{G}}_{k,m}}({{\bf{A}}_m})$ is determined by the antenna orientation. This is because, as the multi-path signals are observed by the 6DMA from different AoAs, the resulting effective aperture and polarization loss factor both vary over
the different 6DMA’s orientations. According to Friis transmission formula, the antenna gain at the $l$-th path is given by~\cite{2015antennatheroy}
\begin{equation}\label{antenna_gain}
	[\mathbf G_{k,m}(\mathbf A_m)]_{l,l} = \frac{\lambda}{4\pi d_{k,m}}\sqrt{\hat g_{k,m}^{l}(\mathbf{A}_m)\bar g_{k,m}^{l}(\mathbf{A}_m)},
\end{equation} 
where ${\lambda^2}/({4\pi d_{k,m}})^2$ represents the free-space path loss with $d_{k,m}$ denoting the distance between AP $m$ and UT $k$. 
$\hat g_{k,m}^{l}(\mathbf{A}_m)$ represents effective aperture loss, which describes how much power is captured from a given plane wave with the same polarization as the receive antenna.
The effective antenna area equals the product of the maximal value of the effective area and the projection of the 6DMA plane normal to the signal direction. Thus, the effective aperture loss is given by~\cite{gainloss}
\begin{equation}
	\hat g_{k,m}^{l}(\mathbf{A}_m) \triangleq \hat g(\mathbf{u}_m;\mathbf{\psi}_{k,m}^l)=\max\{\mathbf{d}(\mathbf{\psi}_{k,m}^l)^{\mathrm{T}}\mathbf{u}_m,0\}.
\end{equation}
When the wave is incident from the back of the antenna, the effective antenna aperture is generally considered to be zero~\cite{2015antennatheroy}. In other words, in that case of $\mathbf{d}(\mathbf{\psi}_{k,m}^l)^{\mathrm{T}}\mathbf{u}_m<0$, we set $\hat g(\mathbf{u}_m;\mathbf{\psi}_{k,m}^l)=0$ \footnote{Although the radiation patterns are varying with different types of antennas, they can all be expressed as the functions of the AoA. Thus, the proposed system model and the corresponding solution can be applied to 6DMA systems with arbitrary antenna radiation patterns}.
$\bar g_{k,m}^{l}(\mathbf{A}_m)$ denotes the polarization loss, which is caused by the mismatch between the polarization directions of the receive antenna and the incident wave. Consequently, the polarization loss can be defined as the squared norm of the inner product of the receive mode polarization vector at the receive antenna and the electric field vector of incident wave. Since the direction of the electric field is always perpendicular to the direction of wave propagation, the electric field vector $\mathbf{e}_{k,m}^l$ from wave direction $\mathbf{\psi}_{k,m}^l$ should satisfy $\mathbf{d}(\mathbf{\psi}_{k,m}^l)^{\mathrm{T}}\mathbf{e}_{k,m}^l=0$. If we set $\mathbf{v}_m$ as the receive mode polarization vector of receive 6DMA, the corresponding polarization loss is given by~\cite{gainloss}
\begin{equation}\label{SP_loss}
	\bar g_{k,m}^{l}(\mathbf{A}_m) \triangleq \bar g(\mathbf{v}_m;\mathbf{e}_{k,m}^l)=|(\mathbf{e}_{k,m}^l)^{\mathrm{T}}\mathbf{v}_m|^2.
\end{equation}
As observed, the antenna orientation controls the effective aperture
loss and polarization loss, and thus determines the amplitude of the complex path coefficient. 

Recalling the field-response channel model in 
(\ref{channel}), we can see that the channel coefficient is jointly determined by the 6DMA's position and orientation. A small change in 6DMA's position or orientation may result in a significant variation in the channels between the corresponding AP and all UTs.
\subsection{Signal Model}
For the uplink data transmission from the UTs to APs, the received signal at AP $m$ is given by
\begin{equation}
	{r_m}= \sum\limits_{k \in {{\mathcal K}}} {{h_{k,m}}({{\bf{q}}_m},{{\bf{A}}_m}){s_k} + {n_m}} ,
\end{equation}
where $s_k$ is the transmit signal of UT $k$ with power $p_k$ and $n_m\sim \mathcal{CN}(0,\sigma^2)$ is the zero-mean additive white Gaussian noise (AWGN) with power $\sigma^2$. Without loss of generality, we assume that the UTs adopt the same maximum transmit power, i.e, $p_k = p$, $k \in \mathcal K$, which can help improve their achievable rates. As such, we mainly focus
on the impact of 6DMAs’positions and orientations on system performance.
Considering the fully centralized signal processing~\cite{Cellfree2020BE}, the signals received at all APs are forwarded
through the fronthaul links to the CPU for decoding, with the combined signal of UT $k$ given by 
\begin{equation}
	\small
	{\hat s_k} = {\bf{w}}_k^{\rm{H}}{{\bf{h}}_k}\left( {{\tilde{\mathbf{q}}},{\tilde{\mathbf{A}}}} \right){s_k} + \sum\limits_{k' \in {{\mathcal K}\backslash }k} {{\bf{w}}_k^{\rm{H}}{{\bf{h}}_{k'}}\left( {{\tilde{\mathbf{q}}},{\tilde{\mathbf{A}}}} \right){s_{k'}}}  + {\bf{w}}_k^{\rm{H}}{{\bf n}_k},
\end{equation}
where ${{\bf{w}}_k} = {[{w_{1,k}},{w_{2,k}}, \ldots ,{w_{M,k}}]^{\rm{T}}}\in\mathbb C^{M}$ is the collective combining vector and ${{\bf{h}}_k}\left( {{\tilde{\mathbf{q}}},{\tilde{\mathbf{A}}}} \right) = {[{h_{1,k}}({{\bf{q}}_1},{{\bf{A}}_1}),{h_{2,k}}({{\bf{q}}_2},{{\bf{A}}_2}), \ldots ,{h_{M,k}}({{\bf{q}}_M},{{\bf{A}}_M})]^{\rm{T}}}\in\mathbb C^{M}$ denotes the collective channel vector of UT $k$. For ease of notation,  we denote the collective channel matrix and receive combining matrix of all UTs as $\mathbf H \left( {{\tilde{\mathbf{q}}},{\tilde{\mathbf{A}}}} \right) = \left[{{\bf{h}}_1}\left( {{\tilde{\mathbf{q}}},{\tilde{\mathbf{A}}}} \right) ,{{\bf{h}}_2}\left( {{\tilde{\mathbf{q}}},{\tilde{\mathbf{A}}}} \right) ,\dots,{{\bf{h}}_K}\left( {{\tilde{\mathbf{q}}},{\tilde{\mathbf{A}}}} \right) \right]$ and $\mathbf W = [{{\bf{w}}_1},{{\bf{w}}_2},\dots,{{\bf{w}}_K}]$, respectively.  
${\bf n}_k = [n_1,n_2,\dots,n_m]\in\mathbb C^{M}$ collects all noise with mean zero and covariance matrix $\sigma^2\mathbf{I}_M$. Then, the achievable rate of UT $k$ is given by
\begin{equation}\label{RATE}
	{R_k}\left( {\tilde{\mathbf{q}}},{\tilde{\mathbf{A}}},{{\bf {w}}_k} \right) = {\log _2}\left( {1 + {\gamma_k}\left( {\tilde{\mathbf{q}}},{\tilde{\mathbf{A}}},{{\bf {w}}_k} \right)} \right),
\end{equation}
where its signal-to-interference-plus-noise ratio (SINR) is given by
\begin{equation}\label{SINR}
	{\gamma_k}\left( {\tilde{\mathbf{q}}},{\tilde{\mathbf{A}}},{{\bf {w}}_k} \right) =  \frac{{{{\left| {{\bf{w}}_k^{\rm{H}}{{\bf{h}}_k}\left( {{\tilde{\mathbf{q}}},{\tilde{\mathbf{A}}}} \right)} \right|}^2}}}{{{{\left| {\sum\limits_{k' \in {{\mathcal K}}\backslash k} {{\bf{w}}_k^{\rm{H}}{{\bf{h}}_{k'}}\left( {{\tilde{\mathbf{q}}},{\tilde{\mathbf{A}}}} \right)} } \right|}^2} + {{\left\| {{{\bf{w}}_k}} \right\|}^2}{\tilde{\sigma}^2}}},
\end{equation}
with $\tilde{\sigma}^2 = {\sigma}^2/p$ denoting the normalized noise power.

In this paper, we aim to maximize the WSR of multiple UTs by joint design of the APV and AOM as well as  the receive combining matrix of all APs. The optimization problem is thus formulated as
\begin{subequations}\label{op}
	\begin{align}
		\max \limits _{\tilde{\mathbf{q}},\tilde{\mathbf{A}},\mathbf W}&~~\sum \limits _{k\in {{\mathcal K}}}\omega_k {R_k}\left( {\tilde{\mathbf{q}}},{\tilde{\mathbf{A}}},{{\bf {w}}_k} \right) \label{opA}\\
		\mbox { s.t.}~~ 
		& {\mathbf{q}}_m \in \mathcal{C}_{m}, \forall m \in \mathcal{M}, \label{opB}\\
		& {\mathbf{A}}_m^{\rm T}{\mathbf{A}}_m = {\bf I}_2, \forall m \in \mathcal{M}, \label{opC}
	\end{align}   
\end{subequations}
where $\omega_k$ is
the  weighting coefficient of UT $k$. 
Constraint (\ref{opB}) confines that each 6DMA can only move in the given receive region. 
Constraint (\ref{opC}) ensures the inherent constraints of two orthogonal unit vectors used to characterize the antenna's orientation. Note that problem (\ref{op}) is a non-convex optimization problem with highly-coupled optimization variables. Consequently, it is challenging to obtain the globally optimal solution. In the following, an AO-based algorithm is developed to obtain a suboptimal solution for problem (\ref{op}).

\section{Proposed Solution}\label{sec_ProposedSolution}
In this section, we first transform the original problem into a more tractable quadratic form by introducing auxiliary variables. Then, for the given collective channel, the closed-form receiving combining is obtained by the MMSE receiver. Next, the APV and AOM are alternately optimized by leveraging the SCA technique and Riemannian manifold optimization, respectively. Finally, the AO of the APV, AOM and the receive combining matrix are implemented to obtain a suboptimal solution for problem (\ref{op}).

\subsection{Problem Transformation}
Inspired by the fractional programming algorithm introduced in \cite{FP2018}, the Lagrangian dual transform and quadratic transform are applied to reformulate the objective function in problem (\ref{op}) as $R(\tilde{\mathbf q},\tilde{\mathbf A},\bf W,\bm \alpha, \bm \beta )$ shown in (\ref {quadratic_form}) at the top of next page, 
where $\mathrm{A}_k = \mathbf{w}_k^{\rm{H}}{{\bf{h}}_k}( {{\tilde{\mathbf{q}}},{\tilde{\mathbf{A}}}})$ and $\mathrm{B}_k = {\left\| {{{\bf{w}}_k}} \right\|^2}{{\tilde \sigma }^2} + \sum\limits_{k' \in {{\mathcal K}}} {{{\left| {{\bf{w}}_k^{\rm{H}}{{\bf{h}}_{k'}( {{\tilde{\mathbf{q}}},{\tilde{\mathbf{A}}}})}} \right|}^2}}$.
$\bm \alpha$ and $\bm \beta$ are introduced as auxiliary variables, which refer to a collection of auxiliary variables $\{\alpha_k\}_{k \in \mathcal K}$ and $\{\beta_k\}_{k \in \mathcal K}$, respectively. When
$(\tilde{\mathbf q},\tilde{\mathbf A},\bf W)$ are fixed, the optimal $\bm \alpha$ and  $\bm \beta$ can be explicitly determined
by setting $\partial R / \partial \alpha_k$ and $\partial R / \partial \beta_k$ to zeros, i.e.,
\setcounter{equation}{14}
\begin{equation}\label{update_alpha}
	{\alpha _k} = \frac{{{{\left| {{\bf{w}}_k^{\rm{H}}{{\bf{h}}_k}} \right|}^2}}}{{\sum\limits_{k' \in {{\mathcal K}}\backslash k} {{{\left| {{\bf{w}}_k^{\rm{H}}{{\bf{h}}_{k'}}} \right|}^2}}  + {{\left\| {{{\bf{w}}_k}} \right\|}^2}{{\tilde \sigma }^2}}},
\end{equation}
\begin{equation}\label{update_beta}	
	{\beta _k} = \frac{{\sqrt {{\omega _k}(1 + {\alpha _k})} {\bf{w}}_k^{\rm{H}}{{\bf{h}}_k}}}{{\sum\limits_{k' \in {{\mathcal K}}} {{{\left| {{\bf{w}}_k^{\rm{H}}{{\bf{h}}_{k'}}} \right|}^2}}  + {{\left\| {{{\bf{w}}_k}} \right\|}^2}{{\tilde \sigma }^2}}},
\end{equation}
where $\mathbf h_k \triangleq {{\bf{h}}_k}\left( {{\tilde{\mathbf{q}}},{\tilde{\mathbf{A}}}} \right)$ with its $m$-th component ${h}_{k,m}\triangleq {{h}_{k,m}}\left( {{\tilde{\mathbf{q}}},{\tilde{\mathbf{A}}}} \right) $ for simplicity of notation, and the following notations for channels will follow this rule unless specified otherwise.
In order to solve problem (\ref{op}) over $(\tilde{\mathbf q},\tilde{\mathbf A},\bf W)$, we can
consider the following equivalent problem over $(\tilde{\mathbf q},\tilde{\mathbf A},\bf W,\bm \alpha, \bm \beta )$:
\begin{subequations}\label{qua}
	\begin{align}
		\max \limits _{\tilde{\mathbf{q}},\tilde{\mathbf{A}},\mathbf W,\bm \alpha, \bm \beta}&~~R(\tilde{\mathbf q},\tilde{\mathbf A},\bf W,\bm \alpha, \bm \beta ) \label{quaA}\\
		\mbox { s.t.}~~ 
		& \text{(\ref{opB})},\text{(\ref{opC})}. \label{qua_b}
	\end{align}   
\end{subequations}
Thus, the original problem can be efficiently solved by the AO of $(\bm \alpha,\bm \beta)$ and $(\tilde{\mathbf q},\tilde{\mathbf A},\bf W)$. Specifically, we can first update $\bm \alpha$ and $\bm \beta$ according
to (\ref{update_alpha}) and (\ref{update_beta}), and then optimize $(\tilde{\mathbf q},\tilde{\mathbf A},\bf W)$ with given $\bm \alpha$ and $\bm \beta$. 
\subsection{Optimization of Receive Combining}
Note that (\ref{SINR}) has a similar form as the generalized Rayleigh quotient~\cite{2007Wireless}. Thus, the optimal receive combining matrix can be derived in closed form based on the MMSE receiver given by  
\begin{equation}
	\label{MMSE}
	\begin{split}
		\hat{\mathbf W}(\tilde{\mathbf q},\tilde{\mathbf A})&=\left(\mathbf H(\tilde{\mathbf q},\tilde{\mathbf A})\mathbf{H}(\tilde{\mathbf q},\tilde{\mathbf A})^{\mathrm H}+\tilde{\sigma}^2\mathbf I_M\right)^{\dagger}\mathbf {H}(\tilde{\mathbf q},\tilde{\mathbf A})\\
		&\triangleq\left[\hat{\mathbf{w}}_1,\hat{\mathbf{w}}_2,\dots,\hat{\mathbf{w}}_K\right],\\   
	\end{split}
\end{equation}
which leads to the maximum SINR for each UT.
It can be seen that the receive combining matrix is determined by the collective channel, which is affected by all antennas' positions and orientations. Next, we will first use the given receive combining matrix to optimize the antennas' positions and orientations, and then update the receive combining matrix via (\ref{MMSE}) in each iteration to match the collective channel.

\subsection{Optimization of Antenna Position}
For given $(\tilde{\mathbf A},\bf W,\bm \alpha, \bm \beta )$, problem (\ref{qua})  is still a highly non-convex problem over $\tilde{\mathbf q}$. To make problem (\ref{qua}) more tractable, an AO strategy is employed, in which only one 6DMA's position is optimized for each iteration with the other 6DMAs' positions being fixed. As such, the part of the objective function in (\ref{quaA}) determined by $\mathbf{q}_m$ can be extracted separately as
\begin{equation}\label{op_q}
	\small
F({\mathbf{q}_m}) \triangleq \sum\limits_{k \in \mathcal K} { {2\Re \{ {h_{k,m}}\left( {{{\bf{q}}_m}} \right){c_{k,m}}\}  - v_{m}{{\left| {{h_{k,m}}\left( {{{\bf{q}}_m}} \right)} \right|}^2}} },
\end{equation}
where ${{h_{k,m}}\left( {{{\bf{q}}_m}} \right)}$ is the channel coefficient function with respect to $\mathbf q_m$ between AP $m$ and UT $k$, which can be obtained using (\ref{channel}). The variables  $c_{k,m}$ and $v_{m}$ are fixed for given $\{\mathbf q_{m'}\}_{m' \in \mathcal M\backslash m}$ and $(\tilde{\mathbf A},\bf W,\bm \alpha, \bm \beta )$, which can be expressed as 
\begin{equation}\label{update_c}
	\begin{aligned}	 
	&{c_{k,m}} = \sqrt {{\omega _k}(1 + {\alpha _k})} \beta _k^*w_{k,m}^ *  \\
	&- \sum\limits_{k' \in \mathcal K} {|{\beta _{k'}}{|^2}w_{k',m}^ * \sum\limits_{m' \in \mathcal M\backslash m} {{w_{k',m'}}h_{k,m'}^ * } },
	\end{aligned} 
\end{equation}
\vspace{-1 em}
\begin{equation}\label{update_v}
	{v_{m}} = \sum\limits_{k \in {\mathcal K}} {|{\beta _k}{|^2}w_{k,m}^ * {w_{k,m}}}.
\end{equation}
Then, the subproblem for optimizing $\mathbf {q}_m$ can be simplified as
\begin{subequations}\label{op:APV}
	\begin{align}	
		\max \limits _{\mathbf{q}_m}&~~F({\mathbf{q}_m}) \label{op:APVA}\\
		\mbox { s.t.}~~ 
		& {\mathbf{q}}_m \in \mathcal{C}_{m}, \label{op:APVB}
	\end{align}   
\end{subequations}
To make problem (\ref{op:APV}) tractable, we adopt the majorization-minimization (MM) method. Specifically, with given local point $\mathbf q_m^i$ in the $i$-th iteration of SCA, the quadratic term in $F(\mathbf q_m)$ can be 
upper-bounded by \cite[Example 13]{MM2017Sun}
\begin{equation}\label{MM}
	\begin{aligned}
		&v_{m}{{\left| {{h_{k,m}}(\mathbf q_m) } \right|}^2}=\mathbf{f}_{k,m}^{\text H}(\mathbf q_m)\mathbf C_{k,m}\mathbf{f}_{k,m}(\mathbf q_m)\\
		&\leq \mathbf{f}_{k,m}^{\text H}(\mathbf q_m)\mathbf \Lambda_{k,m}\mathbf{f}_{k,m}(\mathbf q_m)\\
		&-2\Re\left\{\mathbf{f}_{k,m}^{\text H}(\mathbf q_m)(\mathbf\Lambda_{k,m}
		-\mathbf C_{k,m})\mathbf{f}_{k,m}(\mathbf q_m^i)\right\} \\
		&+\mathbf{f}_{k,m}^{\text H}(\mathbf q_m^i)(\mathbf{\Lambda}_{k,m}-\mathbf{C}_{k,m})\mathbf{f}_{k,m}(\mathbf q_m^i).
	\end{aligned}
\end{equation}
\begin{figure*}[!t]
	\normalsize
	\setcounter{MYtempeqncnt}{\value{equation}}
	\setcounter{equation}{13}
	\begin{equation}
		\label{quadratic_form}
		R({\tilde{\mathbf q},\tilde{\mathbf A},\bf W,\bm \alpha, \bm \beta })=\sum\limits_{k \in {{\mathcal K}}} {{\omega _k}} [\log (1 + {\alpha _k}) - {\alpha _k}]+ \sum\limits_{k \in {{\mathcal K}}} 2 \sqrt {{\omega _k}\left( {1 + {\alpha _k}} \right)} \Re \{ \beta _k^*{{\rm{A}}_k}\}  - |{\beta _k}{|^2}{{\rm{B}}_k}.
	\end{equation}
	\setcounter{MYtempeqncnt}{\value{equation}}
	\hrulefill
	\vspace{-1 em}
\end{figure*}where $\mathbf{C}_{k,m} =v_m {{\bf{G}}_{k,m}}({{\bf{A}}_m}){{\bf{a}}_{k,m}}{{\bf{a}}_{k,m}^{\text H}}{{\bf{G}}_{k,m}^{\text H}}({{\bf{A}}_m})$ and $\mathbf{\Lambda}_{k,m} = \varpi_{k,m}\mathbf I_{L_{k,m}}$ with $\varpi_{k,m}$ denoting the maximum eigenvalue of $\mathbf{C}_{k,m}$. As such, we can combine (\ref{op_q}) and (\ref{MM}) to construct a surrogate function that
locally approximates $F(\mathbf{q}_m)$ shown in (\ref{obj_F}) at the top of next page. 
Note that  $\mathbf{f}_{k,m}^{\text H}(\mathbf q_m)\mathbf{\Lambda}_{k,m}\mathbf{f}_{k,m}(\mathbf q_m)=\varpi_{k,m}L_{k,m}$ is constant for given $\mathbf q_m^i$. Consequently, maximizing $F(\mathbf{q}_m)$ can be simplified to maximizing $\bar F(\mathbf{q}_m)$ that is defined in (\ref{obj_F}).
Although $\bar F(\mathbf{q}_m)$ is still non-concave and non-convex over $\mathbf{q}_m$, it can be locally approximated 
by using its second-order Taylor expansion at $\mathbf{q}_m^i$. Specifically, we have \cite[Lemma 1.2.3]{Convex2018}
\begin{equation}\label{surrogate_function}
	\setcounter{equation}{25}
	\begin{aligned}	
\bar F(\mathbf{q}_m)&\geq \bar F(\mathbf{q}_m^i)+\nabla \bar F(\mathbf{q}_m^i)^\mathrm{T}\left(\mathbf{q}_m-\mathbf{q}_m^i\right)\\
&-\frac{\delta_m}2\left(\mathbf{q}_m-\mathbf{q}_m^i\right)^\mathrm{T}\left(\mathbf{q}_m-\mathbf{q}_m^i\right)\\
& = -\frac{\delta_m}{2}\mathbf{q}_m^\mathrm{T}\mathbf{q}_m + \left(\nabla \bar F(\mathbf{q}_m^i)+\delta_m \mathbf{q}_m^i\right)^\mathrm{T}\mathbf{q}_m\\
& +  \bar F(\mathbf{q}_m^i)-\frac{\delta_m}{2}(\mathbf{q}_m^i)^\mathrm{T}\mathbf{q}_m^i, 
	\end{aligned}
\end{equation}
where $\nabla \bar F(\mathbf{q}_m^i)$ denotes the gradient vector of $\bar F(\mathbf{q}_m)$, and $\delta_m$ is a positive real number satisfying $\delta_m \mathbf{I}_3\succeq \nabla^2 \bar F(\mathbf{q}_m^i)$. The derivation of the gradient vector and Hessian matrix of $\bar F(\mathbf{q}_m)$ over $\mathbf{q}_m$ as well as the closed-form expression of $\delta_m$ are detailed in Appendix \ref{app:A}. Ignoring the constant term of the surrogate function in (\ref{surrogate_function}), problem (\ref{op:APV}) can be equivalently transformed into
\begin{subequations}\label{op:quaAPV}
	\begin{align}
		\max \limits _{\mathbf{q}_m}&~~-\frac{\delta_m}{2}\mathbf{q}_m^\mathrm{T}\mathbf{q}_m + \left(\nabla \bar F(\mathbf{q}_m^i)+\delta_m \mathbf{q}_m^i\right)^\mathrm{T}\mathbf{q}_m\label{op:quaAPVA}\\
		\mbox { s.t.}~~ 
		& \text{(\ref{op:APVB})}.\label{op:quaAPVB}
	\end{align}   
\end{subequations}
Obviously, problem (\ref{op:quaAPV}) is a convex quadratic programming problem over $\mathbf{q}_m$ and its optimal solution is given by
\begin{equation}\label{best_q}
	\mathbf{q}_m^{i+1} = \mathcal{B}\left(\frac{1}{\delta_m}\nabla \bar F(\mathbf{q}_m^i)+\mathbf{q}_m^i\right),
\end{equation}
where $\mathcal{B}\left(\mathbf{q} \right)$ is a projection function. It guarantees the constraint (\ref{op:APVB}) by forcing position components outside the feasible region to the nearest boundary, i.e,
\begin{equation}\label{bound_APV}
	[\mathcal{B}(\mathbf{q})]_u=\left\{
	\begin{aligned}
		&[\mathbf{q}]_u^{\text{min}}, ~~\text{if } [\mathbf{q}]_u<[\mathbf{q}]_u^{\text{min}},\\
		&[\mathbf{q}]_u^{\text{max}},~~\text{if }[\mathbf{q}]_u>[\mathbf{q}]_u^{\text{max}},\\
		&\left[{\mathbf{q}}\right]_{u},~~\text{otherwise},
	\end{aligned}\right.
\end{equation}
where $[\mathbf{q}]_u^{\text{max}}$ and $[\mathbf{q}]_u^{\text{max}}$ are the lower-bound and upper-bound on the feasible region of the $u$-th element of $\mathbf{q}$, respectively.
Hereto, by repeatedly updating $\mathbf{q}_m^i$ via (\ref{best_q}) during the iterations, the objective function of problem (\ref{op:APV}) can converge to a suboptimal value. The details of the proposed SCA-based algorithm for solving problem (\ref{op:APV}) are summarized in Algorithm \ref{al_1}. Specifically, the initialized $\mathbf{q}_m^0$ is given by the input $\mathbf{q}_m^{\mathrm {in}}$. Then, in lines 4-7, we repeatedly update $\mathbf{q}_m^i$ via (\ref{best_q}), where $\nabla \bar F \left(\mathbf{q}_m^{i}\right)$ and $\delta_m$ have closed-formed expressions given by (\ref{delta_F}) and (\ref{delta_m}), respectively.
Finally,  if the increase of the objective function is below the convergence threshold $\epsilon_1$, we output $\mathbf{q}_m^i$ of the last iteration as a suboptimal solution for problem (\ref{op:APV}).
\begin{algorithm}[t] \small
	\caption{SCA-based algorithm for solving problem~(\ref{op:APV}).}
	\label{al_1}
	\begin{algorithmic}[1]
		\REQUIRE ~$\mathbf{q}_m^{\mathrm {in}}$, $\mathbf{H}(\tilde{\mathbf q},\tilde{\mathbf A})$, $\mathbf W$, $\bm \alpha, \bm \beta$, $\{c_{k,m}\}_{k\in\mathcal K} $, $v_m$, $p$, $\sigma^2$, $\lambda$, $\epsilon_1$. 
		\ENSURE ~$\mathbf{q}_m$. \\
		\STATE Initialize iteration index $i=0$ and $\mathbf{q}_m^{0}=\mathbf{q}_m^{\mathrm {in}}$. 
		\REPEAT
		\STATE Compute the surrogate function $\bar F \left(\mathbf{q}_m^{i}\right)$ via (\ref{obj_F}).
		\STATE Calculate $\nabla \bar F \left(\mathbf{q}_m^{i}\right)$ and $\delta_m$ via (\ref{delta_F}) and (\ref{delta_m}).
		\STATE Obtain $\mathbf{q}_m^{i+1}$ according to (\ref{best_q}).
		\STATE Update collective channel matrix via (\ref{channel}) for changing $m$-th 6DMA's position to $\mathbf{q}_m^{i+1}$.
		\STATE Update $i\leftarrow i+1$.	
		\UNTIL $\bar F \left(\mathbf{q}_m^{i}\right)-\bar F \left(\mathbf{q}_m^{i-1}\right)<\epsilon_1$.
		\STATE Set $\mathbf{q}_m = \mathbf{q}_m^{i}$.
		\RETURN $\mathbf{q}_m$.
	\end{algorithmic}
\end{algorithm}

\subsection{Optimization of Antenna Orientation}
Similarly, we optimize each 6DMA's orientation with the others being fixed. Then, for given $\{\mathbf A_{m}\}_{m' \in \mathcal M\backslash m}$ and $\left(\tilde{\mathbf q},\bf W,\bm \alpha, \bm \beta \right)$, 
the part of the objective function in (\ref{quaA}) determined by $\mathbf{A}_m$ can be extracted separately as
\begin{equation}\label{op_A}
	\small
	Q({\mathbf{A}_m})\triangleq \sum\limits_{k \in \mathcal K} { {2\Re \{ {h_{k,m}}\left( {{{\bf{A}}_m}} \right){c_{k,m}}\}  - v_{m}{{\left| {{h_{k,m}}\left( {{{\bf{A}}_m}} \right)} \right|}^2}} },
\end{equation}
where ${{h_{k,m}}\left( {{{\bf{A}}_m}} \right)}$ is the channel coefficient function with respect to $\mathbf A_m$ between AP $m$ and UT $k$, which can also be obtained using (\ref{channel}). The variables $c_{k,m}$ and $v_{m}$ can be similarly obtained via (\ref{update_c}) and (\ref{update_v}) for given $\{\mathbf A_{m'}\}_{m' \in \mathcal M\backslash m}$ and $(\tilde{\mathbf q},\bf W,\bm \alpha, \bm \beta )$, respectively. 
As such, the subproblem for optimizing $\mathbf {A}_m$ can be formulated as
\begin{subequations}\label{op:AOM}
	\begin{align}
		\max \limits _{\mathbf{A}_m}&~~Q({\mathbf{A}_m})\label{op:AO6DMA}\\
		\mbox { s.t.}~~ 
		& {\mathbf{A}}_m^{\rm T}{\mathbf{A}}_m = {\bf I}_2. \label{op:AOMB}
	\end{align}  
\end{subequations}
It is challenging to solve problem (\ref{op:AOM}) directly due to the non-concave objective function (\ref{op:AO6DMA}) over the non-convex orthogonal matrix constraint (\ref{op:AOMB}). Note that constraint (\ref{op:AOMB}) is consistent with the definition of the Stiefel manifold~\cite{Stiefel2020}, which is characterized by
\begin{equation}\label{stiefel_manifold}
	{{\mathcal S}} = \left\{ {{\bf{A}} \in {\mathbb R^{3 \times 2}}|{{\bf{A}}^{\rm{T}}}{\bf{A}} = {{\bf{I}}_2}} \right\}.
\end{equation}
\begin{figure*}[!t]
	\normalsize
	\setcounter{MYtempeqncnt}{\value{equation}}
	\setcounter{equation}{23}
	\begin{equation}\label{obj_F}
		\begin{aligned}	
			F(\mathbf q_m) &\geq \underbrace{\sum\limits_{k \in {\mathcal K}}
				2\Re\left\{\mathbf{f}_{k,m}^{\text H}(\mathbf q_m)\left[(\mathbf{\Lambda}_{k,m}-\mathbf{C}_{k,m})\mathbf{f}_{k,m}(\mathbf q_m^i)+{c}_{k,m}{{\bf{G}}_{k,m}}({{\bf{A}}_m}){{\bf{a}}_{k,m}}\right]\right\}}_{\triangleq\bar F(\mathbf q_m)}\\
			&\underbrace{-\mathbf{f}_{k,m}^{\text H}(\mathbf q_m)\mathbf{\Lambda}_{k,m}\mathbf{f}_{k,m}(\mathbf q_m)
				-\mathbf{f}_{k,m}^{\text H}(\mathbf q_m^i)(\mathbf{\Lambda}_{k,m}-\mathbf{C}_{k,m})\mathbf{f}_{k,m}(\mathbf q_m^i)}_{\text {constant}}.
		\end{aligned}
	\end{equation}
	\setcounter{equation}{\value{MYtempeqncnt}}
	\hrulefill
	\vspace{-1 em}
\end{figure*}It means that the optimization of $\mathbf A_m$ over the Stiefel manifold inherently satisfies constraint (\ref{op:AOMB}).
In order to solve problem (\ref{op:AOM}) efficiently, we propose a Riemannian manifold optimization-based algorithm, which converts constrained optimization problems into non-constrained optimization problems on manifolds and find solutions that naturally satisfy the original constraints.

First, we introduce some definitions and terminologies in Riemannian manifold optimization~\cite{Reman2021}. The \emph{tangent space} $T_{\mathbf A}\mathcal S$ of the Stiefel manifold at point $\mathbf A$ is composed of all tangent matrices of the Stiefel manifold at point $\mathbf A$, which is denoted as
\begin{equation}\label{tangent_space}
{T_{{{\bf{A}}}}}{\mathcal S}  = \left\{ {{\bf{Z}} \in {\mathbb R^{3 \times 2}}|{{\bf{Z}}^{\rm{T}}}{{\bf{A}}} + {\bf{A}}^{\rm{T}}{\bf{Z}} = \mathbf{0}_2} \right\},
\end{equation}
where $\bf Z$ represents any tangent matrix at the point $\bf A$ on the Stiefel manifold.
Among all tangent matrices, one of them which yields the greatest increase of the objective function is defined as \emph{Riemanian gradient}. The Riemannian gradient of function $Q$ at $\bf A$ is given by the orthogonal projection of the Jacobian matrix $\nabla Q_{\mathbf A}$ onto the tangent space ${T_{{{\bf{A}}}}}{\mathcal S}$, i.e.,
\begin{equation}\label{Riemanian_gradient}
	\text{grad}_{\mathbf A}Q = \nabla_{\mathbf A} Q-\mathbf A
	\frac{\mathbf A^{\mathrm{T}}\nabla_{\mathbf A} Q+\left(\nabla_{\mathbf A} Q\right)^{\mathrm{T}}\mathbf A}{2}.
\end{equation}
The Jacobian matrix $\nabla_{\mathbf A} Q$ can be calculated by its numerical definition, i.e.,
\begin{equation}\label{Euclidean_gradient}
	\small
	\begin{aligned}
		&\left[\nabla_{\mathbf A}Q\left({\mathbf{A}}^{n}\right)\right]_{u,v}=\left.\frac{\partial Q\left({\mathbf{A}}\right)}{\partial\left[{\mathbf{A}}\right]_{u,v}}\right|_{{\mathbf{A}}={\mathbf{A}}^{n}}\\
		&=\lim_{\delta\to0}\frac{Q\left({\mathbf{A}}^{n}+\delta\mathbf{E}^{u,v}\right)-Q\left({\mathbf{A}}^{n}\right)}{\delta},1\leq u\leq3,1\leq v\leq2,
	\end{aligned}
\end{equation}
where $n$ represents the iteration index, and $\mathbf{E}^{u,v}\in \mathbb{R}^{3\times 2}$ denotes a matrix with the $u$-th row and $v$-th column element as $1$ and zeros elsewhere.

Then, with the aid of the Riemannian gradient, the conventional optimization methods widely employed in Euclidean space can be exploited to solve this manifold optimization problem. In this paper, we introduce the conjugate gradient method as an effective approach, which updates the search direction for maximizing the objective function in Euclidean space as follows:
\begin{equation}\label{CG}
	{\bm \mu}^{n+1} = \nabla_{\mathbf A}Q\left({\mathbf{A}}^{n+1}\right)+\kappa^n{\bm \mu}^{n},
\end{equation}
where ${\bm \mu}^{n}$ is the search direction at $\mathbf A^{n}$ and $\kappa^n$ is the Ploack-Ribiere parameter to guarantee the convergence of objective function~\cite{NP1999}. Since ${\bm \mu}^{n}$  in tangent space $T_{\mathbf A^n}{\mathcal S}$ and $\nabla_{\mathbf A}Q\left({\mathbf{A}}^{n}\right)$ in tangent space $T_{\mathbf A^{n+1}}{\mathcal S}$ cannot be combined directly, we introduce a mapping between two tangent matrices in different tangent spaces called $\emph{transport}$.  Specifically, the transport for Stiefel manifold is given by~\cite{Stiefel2020} 
\begin{equation}\label{transport}
	\begin{aligned}
		&\text{Trans}({\bm \mu}^{n}) \triangleq T_{\mathbf A^n}{\mathcal S}\mapsto T_{\mathbf A^{n+1}}{\mathcal S}:\\
		& {\bm \mu}^{n}\mapsto {\bm \mu}^{n}-\mathbf A^{n+1}
		\frac{(\mathbf A^{n+1})^{\mathrm{T}}{\bm \mu}^{n}+\left({\bm \mu}^{n}\right)^{\mathrm{T}}\mathbf A^{n+1}}{2}.
	\end{aligned}
\end{equation}
Thus, the update rule of search direction of the Riemannian gradient can be similarly obtained as  
\begin{equation}\label{update_mu}
	{\bm \mu}^{n+1} = \text{grad}_{\mathbf A^{n+1}}Q+\kappa^n\text{Trans}({\bm \mu}^{n}).
\end{equation}

Next, in order to determine the destination on the Stiefel manifold, we introduce the $\emph{retraction}$ operation, which maps a matrix in the tangent space onto the Stiefel manifold. Specifically, for the point $\mathbf A^n$ on manifold $\mathcal S$, the retraction of a tangent matrix $\tau^n{\bm \mu}^{n}$ with search step size $\tau^n$ and search direction ${\bm \mu}^{n}$ can be specified as~\cite{Stiefel2020}
\begin{equation}\label{retraction}
	\begin{aligned}
		\text{Retr}(\tau^n{\bm \mu}^{n}) &\triangleq T_{\mathbf A^n}{\mathcal S}\mapsto {\mathcal S}:\\
	& \tau^n{\bm \mu}^{n}\mapsto \mathcal Q\mathcal R\left({\mathbf A}^{n}+\tau^n{\bm \mu}^{n}\right),
	\end{aligned}
\end{equation}
where $\mathcal Q\mathcal R(\cdot)$ denotes the orthonormal matrix 
in QR decomposition, which decomposes a matrix into an orthonormal matrix and an upper triangular matrix. It can be seen that $\text{Retr}(\mathbf 0)={\mathbf A}^{n}$.
As such, $\mathbf{A}^n$ will be repeatedly updated on the Stiefel manifold, which guarantees the objective function to be non-decreasing over $n$.

In summary, we present the complete process of solving problem (\ref{op:AOM}) based on Riemannian manifold optimization in Algorithm \ref{al_2}. The $m$-th 6DMA's orientation is first initialized with the input $\mathbf{A}_m^{\mathrm{in}}$,
and we then calculate its Riemannian gradient as the initial search direction. In lines 3-10, the $m$-th 6DMA's orientation is optimized based on the conjugate gradient method, such that the objection function in (\ref{op:AO6DMA}) keeps monotonically
non-decreasing after each iteration. The algorithm terminates if the increase of the objective function is below a predefined threshold $\epsilon_2$, and finally converges to a suboptimal solution of problem (\ref{op:AOM}).
\begin{algorithm}[t] \small
	\caption{Riemanian manifold optimization-based algorithm for solving problem~(\ref{op:AOM}).}
	\label{al_2}
	\begin{algorithmic}[1]
		\REQUIRE ~$\mathbf{A}_m^{\mathrm{in}}$, $\mathbf{H}(\tilde{\mathbf q}, \tilde{\mathbf A})$, $\mathbf W$, $\bm \alpha, \bm \beta$, $\{c_{k,m}\}_{k\in\mathcal K} $, $v_m$, $p$, $\sigma^2$, $\lambda$, $\epsilon_2$. 
		\ENSURE ~$\mathbf{A}_m$. \\
		\STATE Initialize iteration index $n=0$ and $\mathbf{A}_m^{0}=\mathbf{A}_m^{\mathrm{in}}$.
		\STATE Calculate the initial search direction $\bm \mu^{0} = \text{grad}_{\mathbf{A}_m^{0} }Q$ via (\ref{Riemanian_gradient}).
		\REPEAT
		\STATE Choose the step size $\tau^{n}$ by Armijo backtracking line search.
		\STATE Update $\mathbf{A}_m^{n+1}$ by retraction in (\ref{retraction}).
		\STATE Calculate Riemanian gradient $\text{grad}_{\mathbf A_m^{n+1}}Q$ via (\ref{Riemanian_gradient}).
		\STATE Calculate the vector tranport $\text{Trans}({\bm \mu}^{n})$ via (\ref{transport}) and choose Polak-Ribiere parameter $\kappa ^n$ via  backtracking line search.
		\STATE Update conjugate search direction $\bm{\mu}^{n+1}$ via (\ref{update_mu}).
		\STATE Update $n\leftarrow n+1$.	
		\UNTIL $Q \left(\mathbf{A}_m^{n}\right)-Q \left(\mathbf{A}_m^{n-1}\right)<\epsilon_2$.
		\STATE Set $\mathbf{A}_m = \mathbf{A}_m^{n}$.
		\RETURN $\mathbf{A}_m$.
	\end{algorithmic}
\end{algorithm}

\subsection{Overall Algorithm}\label{Sec:overall_algorithm}
The overall algorithm for alternately optimizing the APV
and AOM as well as receive combining matrix is summarized in Algorithm $\ref{al_3}$. In line 1, each 6DMA's position is initialized randomly within the moving region and its orientation is initialized with the normal vector pointing to one random UT and the receive mode polarization vector paralleling to its electric field vector. Then, in lines 4-10, for given $(\tilde{\mathbf{A}}^{t},{\mathbf{W}}^{t})$, each 6DMA's position is alternately optimized by using Algorithm \ref{al_1}. In addition, the MMSE-based receive combining matrix and auxiliary variables should be accordingly updated to match the channel variations caused by antenna repositioning. Next, in lines 11-16, for given $(\tilde{\mathbf{q}}^{t+1},{\mathbf{W}}^{t})$, each 6DMA's orientation is alternately optimized by using Algorithm \ref{al_2}. Similarly, the receive combining matrix and auxiliary variables are updated as the antenna orientation changes. Thus, in lines 3-19, the AO of APV and AOM as well as receive combining matrix are iteratively executed until the increase of objective function in (\ref{opA}) is below a predefined threshold $\epsilon_3$, which finally yields the local optimum for problem (\ref{op}).
\begin{algorithm}[t] \small
	\caption{The overall algorithm for solving problem~(\ref{op}).}
	\label{al_3}
	\begin{algorithmic}[1]
		\REQUIRE ~$M$, $K$, $\{L_{k,m}\}$, $\{\theta_{k,m}^l\}$, $\{\phi_{k,m}^l\}$, $\{\mathbf e_{k,m}^l\}$, $\{\mathbf a_{k,m}\}$, $p$, $\sigma^2$, $\lambda$, $\epsilon_3$.
		\ENSURE ~$\tilde{\mathbf{q}}^\star$,$\tilde{\mathbf{A}}^\star$,$\mathbf{W}^\star$. \\
		\STATE Initialize iteration index $t=0$, $\tilde{\mathbf{q}}^0$ and $\tilde{\mathbf{A}^0}$. 
		\STATE  Calculate the initial collective channel matrix $\mathbf{H}(\tilde{\mathbf q}^0,\tilde{\mathbf A}^0)$ and receive combining matrix $\mathbf{W}(\tilde{\mathbf q}^0,\tilde{\mathbf A}^0)$.
		\REPEAT
		
		\FOR {$m = 1$ to $M$}
		\STATE Obtain $\{c_{k,m}\}_{k\in\mathcal K} $ and $v_m$ via (\ref{update_c}) and (\ref{update_v}).
		\STATE Update $\mathbf{q}_m^{t+1}$ by solving problem (\ref{op:APV}) via Algorithm \ref{al_1} for given input $\mathbf{q}_m^{t}$.
		\STATE Calculate the FRVs $\{{\bf{f}}_{k,m}({\bf{q}}_m)\}_{k \in \mathcal K}$ via (\ref{FRV}) and the resulting collective channel matrix. 
		\STATE Obtain the receive combining matrix via (\ref{MMSE}).
		\STATE Update auxiliary variables $\bm \alpha$ and $\bm \beta$ via (\ref{update_alpha}) and (\ref{update_beta}).
		\ENDFOR
		
		\FOR {$m = 1$ to $M$}
		\STATE Obtain $\{c_{k,m}\}_{k\in\mathcal K} $ and $v_m$ via (\ref{update_c}) and (\ref{update_v}).
		\STATE Update $\mathbf{A}_m^{t+1}$ by solving problem (\ref{op:AOM}) via Algorithm \ref{al_2} for given input $\mathbf{A}_m^{t}$.
		\STATE Calculate the effective antenna gain matrices $\{{\bf{G}}_{k,m}({\bf{A}}_m)\}_{k \in \mathcal K}$ via (\ref{antenna_gain}) and the resulting collective channel matrix. 
		\STATE Obtain the receive combining matrix via (\ref{MMSE}).
		\STATE Update auxiliary variables $\bm \alpha$ and $\bm \beta$ via (\ref{update_alpha}) and (\ref{update_beta}).
		\ENDFOR
		\STATE Calculate the collective channel matrix $\mathbf H(\tilde{\mathbf{q}}^{t},\tilde{\mathbf{A}}^{t})$ and the receive combining matrix ${\mathbf W}^{t+1}$.
		\STATE Update $t\leftarrow t+1$.
		\UNTIL $R \left(\tilde{\mathbf{q}}^{t},\tilde{\mathbf{A}}^{t},{\mathbf W}^t\right)-R \left(\tilde{\mathbf{q}}^{t-1},\tilde{\mathbf{A}}^{t-1},{\mathbf W}^{t-1}\right)<\epsilon_3$.
		\STATE Set $\tilde{\mathbf{q}}^\star = \tilde{\mathbf{q}}^t$, $\tilde{\mathbf{A}}^\star = \tilde{\mathbf{A}}^t$ and $\mathbf{W}^\star =\mathbf{W}^t$.
		\RETURN $\tilde{\mathbf{q}}^\star$,$\tilde{\mathbf{A}}^\star$,$\mathbf{W}^\star$.
	\end{algorithmic}
\end{algorithm}

The convergence and computational complexity of Algorithm \ref{al_3} are analyzed
as follows.
Since Algorithm \ref{al_1} and Algorithm \ref{al_2}  are both gradient-based algorithms, the objective functions in (\ref{op:APVA}) and (\ref{op:AO6DMA}) are non-decreasing and upper-bounded after each iteration, thereby guaranteeing the convergences of Algorithm \ref{al_1} and Algorithm \ref{al_2}. In addition, Algorithm \ref{al_3} is an AO-based algorithm,
whose convergence is guaranteed by the following inequality:
\begin{equation}
	\label{convergence_inner}
	\begin{split}
		R \left(\tilde{\mathbf{q}}^{t},\tilde{\mathbf{A}}^{t},{\mathbf W}^t\right)
		&\overset{(a)}{\ge} 
		R \left(\tilde{\mathbf{q}}^{t},\tilde{\mathbf{A}}^{t},{\mathbf W}^{t-1}\right)\\
		&\overset{(b)}{\ge} 
		R \left(\tilde{\mathbf{q}}^{t},\tilde{\mathbf{A}}^{t-1},{\mathbf W}^{t-1}\right)\\
		&\overset{(c)}{\ge}
		R \left(\tilde{\mathbf{q}}^{t-1},\tilde{\mathbf{A}}^{t-1},{\mathbf W}^{t-1}\right),
	\end{split}
\end{equation}
where $(a)$ holds because ${\mathbf W}^t$ is the optimal MMSE combining matrix for maximizing the SINR of each UT under the current collective channel matrix $\mathbf H(\tilde{\mathbf{q}}^{t},\tilde{\mathbf{A}}^{t})$, and $(b)$ and $(c)$ are guaranteed by the convergence of Algorithm \ref{al_2} and Algorithm \ref{al_1}, respectively. It means that the objective value is non-decreasing over iterations in Algorithm~\ref{al_3}.
Meanwhile, the objective value of problem (\ref{op}) is always upper-bounded. Thus, the convergence of the overall algorithm is guaranteed. Moreover, the convergence performance will also be validated by simulation in Section~~\ref{sec_SimulationResults}.

The major computational complexity of optimizing the 6DMA's position in problem (\ref{op:APV}) is of order $\mathcal{O}(L_{\text{max}}^{3.5}+I_{\text{max}}KL_{\text{max}})$ in terms of the maximum number of paths $L_{\text{max}}$, the number of UTs $K$, and the maximum number of iterations $I_{\text{max}}$ in Algorithm~\ref{al_1}, which is determined by the complexity of calculating the maximum eigenvalues of $\mathbf C_{k,m}$ and $\nabla \bar F(\mathbf{q}_m)$.
The major computational complexity of solving the 6DMA's orientation in problem (\ref{op:AOM}) is of order $\mathcal{O}(N_{\text{max}}KL_{\text{max}})$ with the maximum number of iterations $N_{\text{max}}$ in Algorithm~\ref{al_2}. As such, the complexity order of optimizing APV and AOM in lines 4-10 and lines 11-17 is  $\mathcal{O}(M(L_{\text{max}}^{3.5}+I_{\text{max}}KL_{\text{max}}+N_{\text{max}}KL_{\text{max}}))$. In addition, the complexity order of designing the receiving combining matrix via (\ref{MMSE}) is $\mathcal{O}(M^3)$.
As a result, with the maximum number $T_{\text{max}}$ of iterations in Algorithm~\ref{al_3}, the total computational complexity of the overall algorithm is of order $\mathcal{O}(T_{\text{max}}M(M^3+L_{\text{max}}^{3.5}+I_{\text{max}}KL_{\text{max}}+N_{\text{max}}KL_{\text{max}}))$.

\subsection{Offline Solution Based on Statistical CSI}\label{Sec:Offline}
Note that the proposed solution in Algorithm \ref{al_3} requires the accurate instantaneous CSI (including AoAs, PRVs and polarization vectors of all channel paths), which may result in high overhead in channel estimation and antenna movement in practice if the UTs move or the propagation environment changes frequently. To address this issue, we devise an offline optimization algorithm to maximize the average WSR of 6DMA-enhanced multi-AP coordination system based on the statistical CSI, similarly as~\cite{6DMA20204model}. As such, the offline-designed 6DMAs' positions and orientations can be reconfigured over a long time period. Then, problem (\ref{op}) is recast to
\begin{subequations}\label{op_offline}
	\begin{align}
		\max\limits _{\tilde{\mathbf{q}},\tilde{\mathbf{A}}}~~&\mathbb{E}\left\{\max \limits _{{\bf {W}} }~{{R}\left( {\tilde{\mathbf{q}}},{\tilde{\mathbf{A}}},{{\bf {W}} } \right)} \right\}\label{op_offlineA}\\
		\mbox { s.t.}~~ 
		& \text{(\ref{opB})},\text{(\ref{opC})}. \label{op_offlineB}
	\end{align}   
\end{subequations}
In particular, the expectation in (\ref{op_offlineA}) is executed over random channels between the UTs and APs based on a given distribution. In general, it is challenging to derive the expected WSR in closed form. Nonetheless, it can be approximated by the Monte Carlo method~\cite{MC1999}, which involves numerous realizations of random independent instantaneous channels, i.e.,
\begin{equation}
	\mathbb{E}\left\{\max \limits _{{\bf {W}} }~{{R}\left( {\tilde{\mathbf{q}}},{\tilde{\mathbf{A}}},{{\bf {W}}} \right)} \right\} \approx \frac{1}{|\mathcal L|} \sum_{\iota \in \mathcal L}\max \limits _{{\bf {W}}_\iota }{R}_{\iota}\left( {\tilde{\mathbf{q}}},{\tilde{\mathbf{A}}},{{\bf {W}}_\iota } \right),
\end{equation}
where $\mathcal L$ is the set of Monte Carlo simulations, in which each element represents one realization of instantaneous channels, and ${R}_{\iota}\left( {\tilde{\mathbf{q}}},{\tilde{\mathbf{A}}},{{\bf {W}}_\iota } \right) $ can be calculated according to (\ref{RATE}) for the $\iota$-th realization. 
It should be emphasized that the 
APV and AOM are offline-optimized based on statistical CSI and online-applied over instantaneous channels, together with the receive combining.

Note that problem (\ref{op_offline}) has a similar form of the objective function form and the same constraints as problem (\ref{op}). Thus, for the different realizations of instantaneous channels, the optimal receive combining matrix can be obtained by the MMSE receiver according to (\ref{MMSE}). Then, given  $\{\mathbf W_\iota\}_{\iota \in \mathcal L}$, the APV and AOM can still be optimized by the SCA method and Riemannian manifold optimization, respectively, where the SCA and gradient should be averaged over $|\mathcal L|$ channel realizations. As such, 
problem (\ref{op_offline}) can be solved by alternately optimizing $\tilde{\mathbf{q}}$ and $\tilde{\mathbf{A}}$ similarly to that in Algorithm \ref{al_3}. After completing the optimization of APV and AOM, the 6DMA’s positions and orientations will be adjusted accordingly and remain constant for a long period, which significantly reduces the overhead for antenna movement.
Moreover, although the instantaneous channel changes, the deployed 6DMAs can still attain a good system performance for the proposed 6DMA-enhanced multi-AP coordination system, as will be validated by the simulation results in Section \ref{sec_SimulationResults}.

\subsection{Extension to Dual-polarized Antenna}\label{Sec:dual}
 Considering the advantages of dual-polarized antennas in avoiding polarization mismatch between the transceivers, 
 we extend the proposed 6DMA-enhanced multi-AP coordination system from uni-polarized to dual-polarized modes, in which each antenna at the AP has two co-located orthogonal polarizations  and thus can be equivalently regarded as two independent antennas~\cite{2015antennatheroy}.
 
Denoting the vertical and horizontal polarization vectors of the $m$-th dual-polarized antenna as $\mathbf{v}_m^1$ and $\mathbf{v}_m^2$, respectively, we define the collective dual-polarized AOM as $\tilde{\mathbf A}'= [{\mathbf A}'_1,{\mathbf A}'_2,\dots,{\mathbf A}'_M]\in \mathbb{R}^{3\times 3M}$ with $\mathbf{A}'_m = [\mathbf{u}_m,\mathbf{v}_m^1,\mathbf{v}_m^2]$. As such,
the collective channel matrix can be updated as $\mathbf{H}_\text{dual}({\tilde{\mathbf{q}},\tilde{\mathbf{A}}'})= [\mathbf{H}^1;\mathbf{H}^2]\in\mathbb C^{2M \times K} $, where $[\mathbf{H}^1]_{k,m} ={\bf{f}}_{k,m}^{\rm{H}}({{\bf{q}}_m}){{\bf{G}}_{k,m}}\left([\mathbf{u}_m,\mathbf{v}^1_m]\right){{\bf{a}}_{k,m}}$ and $[\mathbf{H}^2]_{k,m} ={\bf{f}}_{k,m}^{\rm{H}}({{\bf{q}}_m}){{\bf{G}}_{k,m}}\left([\mathbf{u}_m,\mathbf{v}^2_m]\right){{\bf{a}}_{k,m}}$.

Then, the maximization of WSR in the 6DMA-enhanced multi-AP coordination system can be similarly formulated as 
\begin{subequations}\label{op_dual}
	\begin{align}
		\max \limits _{\tilde{\mathbf{q}},\tilde{\mathbf{A}}',\mathbf W}&~~\sum \limits _{k\in {{\mathcal K}}}\omega_k {R_k}\left( {\tilde{\bf q}},{\tilde{\bf A}}',{{\bf {w}}_k} \right) \label{op_dualA}\\
		\mbox { s.t.}~~ 
		& {\mathbf{q}}_m \in \mathcal{C}_{m}, \forall m \in \mathcal{M}, \label{op_dualB}\\
		& (\mathbf{A}'_m)^{\mathrm{T}}\mathbf{A}'_m = \mathbf{I}_3, \forall m \in \mathcal{M}. \label{op_dualC}
	\end{align}   
\end{subequations}
Since $\mathbf{A}'_m$ is still on the Stiefel manifold, the dual-polarized antennas' orientations can also be optimized by the proposed Riemannian manifold optimization-based algorithm. Consequently, problem (\ref{op_dual}) can be similarly solved by the gradient ascent-based algorithm. Moreover, in the following simulation section, we will validate
the applicability of the proposed algorithms in the case of dual-polarized antennas, and analyze the impact of dual polarizations on 6DMA-enhanced multi-AP coordination system performance.
\section{Simulation Results}\label{sec_SimulationResults}
In this section, simulation results are presented to evaluate the performance of the proposed 6DMA-enhanced multi-AP coordination system and demonstrate the performance superiority of the proposed algorithms for maximizing the WSR of multiple UTs over various benchmark schemes.
\subsection{Simulation Setup}
\begin{figure}[t]
	\begin{center}
		\includegraphics[width=\figwidth cm]{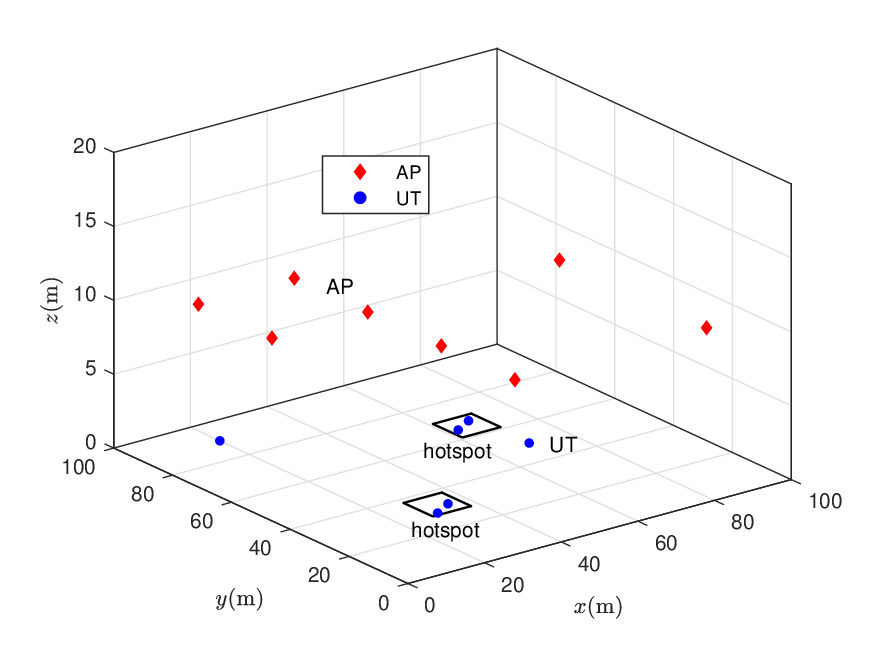}
		\caption{Geographical distribution of APs and UTs in GCCS.}
		\label{fig:Scenario}
	\end{center}
\end{figure}
In the simulations, a typical wireless network is considered where $K$ FA-UTs are served by the $M$ 6DMA-APs. In particular, we consider a 3D multi-hotspot scenario where the majority of UTs are randomly distributed in two non-overlapping hotspot sub-regions and the rest of UTs are uniformly distributed in the remaining areas. To be specific, we develop a 3D global Cartesian coordinate system (GCCS) shown in Fig.~\ref{fig:Scenario} to illustrate the geographical distributions of APs and UTs.
The weighting coefficients for the UTs are set to
ones, i.e., $\omega_k = 1, \forall k \in \mathcal K$.
The 6DMA's moving region at the each AP is set as a cube of size $2\lambda \times 2\lambda \times 2\lambda$. We adopt the geometry Rician fading channel model with Rician factor as $\chi$, in which the numbers of receive paths for all users are assumed to be identical, i.e., $L_{k,m}=L, \forall m \in \mathcal{M}, k \in \mathcal{K} $~\cite{MUMA2023ZL,Conf2023Pi}. Without loss of  generality, we set the first path as the line-of-sight (LoS) path while the rest of paths as non-LoS (NLoS) paths. 
Then, each element of the PRV is assumed to be an independent and identically distributed (i.i.d.) CSCG random variable, 
i.e., $a_{k,m}^1 \sim \mathcal{CN}(0,\frac{\chi}{1+\chi})$ and $a_{k,m}^l \sim \mathcal{CN}(0,\frac{1}{(L-1)(\chi+1)})$ for $m\in\mathcal M, k\in \mathcal K$, and $2 \le l \le L$. 
The elevation and azimuth AoAs of NLoS paths for each UT are generated by environmental scatterers, which are randomly distributed in a cuboid of the size  $100\text{m}\times100\text{m}\times20\text{m}$,
and the AoAs of LoS paths are determined by the geographical locations of UTs and APs. The detailed settings of simulation parameters are listed in Table~\ref{tab:para}, unless specified otherwise. 
Each point in the simulation figures is the average result over $100$ random UT location and channel realizations.
\begin{table}[t]
 	\renewcommand{\arraystretch}{1.3}
	\caption{Simulation Parameters}\label{tab:para}
	\footnotesize
	\begin{center}
		\begin{tabular}{|c|l|c|}
			\hline
			\textbf{Parameter}                        &\multicolumn{1}{c|}{\textbf{Description}}                                       & \textbf{Value} \\
			\hline
			$M$                                 & Number of APs                                   	& $8$ \\
			\hline
			$K$                                 & Number of UTs                                   & $6$ \\
			\hline
			$L$                                 &Number of channel paths for each UT                           		& $5$ \\
			\hline
			$f_c$                       	&Carrier frequency                           	& $2.4$ GHz \\
			\hline
			$\lambda$                       	&Carrier wavelength                           	& $0.125$ m \\
			\hline
			$A$                       &Length of the sides of receive region                                  & $2\lambda$ \\
			\hline
			$\chi$                    & Rician factor                          		& $10$ \\
			\hline
			$\sigma^2$                    & Average noise power                           		& $-80$ dBm \\
			\hline
			$p$                    & Transmit power for each UT                          		& $10$ dBm \\
			\hline
			$\epsilon_1$                    & Convergence threshold in Algorithm~\ref{al_1}   			& $10^{-3}$ \\
			\hline
			$\epsilon_2$                   & Convergence threshold in Algorithm~\ref{al_2}   					& $10^{-3}$ \\
			\hline
			$\epsilon_3$                    & Convergence threshold in Algorithm~\ref{al_3}                 & $10^{-2}$ \\
			\hline
		\end{tabular}
	\end{center}
	\vspace{-0.2 in}
\end{table}
\subsection{Algorithm Convergence Evaluation}

\begin{figure}[t]
	\begin{center}
		\includegraphics[width=\figwidth cm]{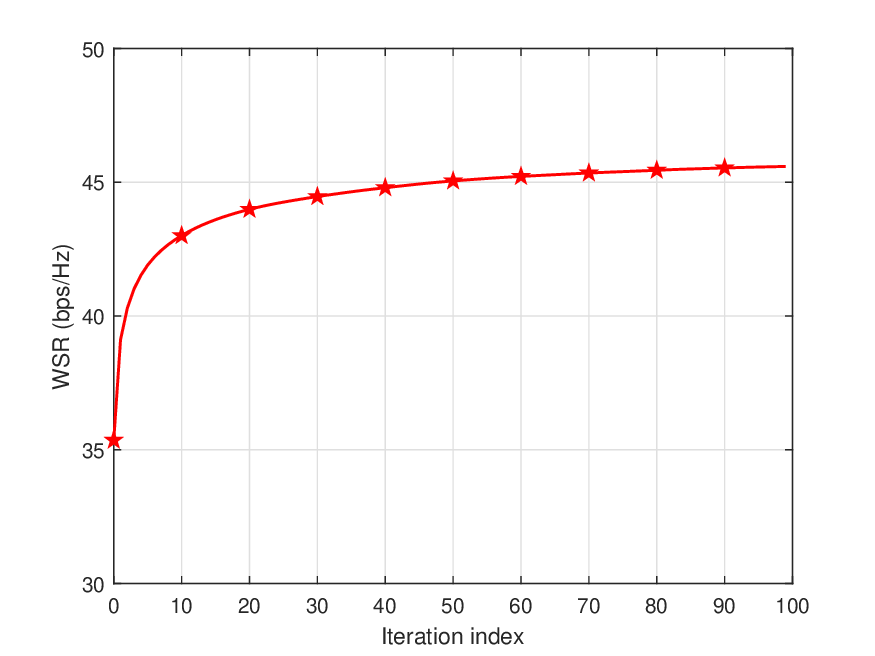}
		\caption{Convergence performance of Algorithms \ref{al_3}.}
		\label{fig:Convergence}
	\end{center}
\end{figure}

In Fig.~\ref{fig:Convergence}, we show the convergence of the proposed AO-based algorithm for the 6DMA-enhanced multi-AP coordination system. 
As can be observed, the WSR in problem (\ref{op}) converges within $100$ iterations, which demonstrates the efficient convergence of the proposed Algorithm~\ref{al_3}.
Moreover, the WSR of all UTs increases from $35.3$ bits per second per Hertz (bps/Hz) to $45.6$ bps/Hz, which means the 6DMA scheme yields about $30\%$ performance improvement, which confirms the effectiveness of our proposed algorithm for obtaining high-quality solutions and improving network communication performance.

\subsection{Performance Comparison with Benchmark Schemes}
The proposed algorithm for 6DMA-enhanced multi-AP coordination systems based on instantaneous CSI and statistical CSI are labeled by ``6DMA" and ``offline-6DMA", respectively.
Besides, we consider four benchmark schemes for performance comparison, namely ``ES", ``6DMA-position'', ``6DMA-orientation'' and ``FA", respectively. 
\begin{itemize}
		
	\item {ES: We uniformly generate $D_p=13^3$ discrete positions and $D_r=12^3$ discrete	orientations for each AP.
		As such, the dimension of search space is $(D_pD_r)^M$, which leads to a prohibitively high computational complexity for the exhaustive search (ES) over all possible solutions. To make the search implementable, we adopt the alternating  selection method to alternately select each 6DMA's position and orientation with the others being fixed. }
	
	\item {6DMA-position: This scheme implements 6DMA with position optimization only at each AP with fixed orientation, i.e., $\mathbf u_m=[1,0,0]^\text{T}$ and $\mathbf v_m=[0,1,0]^\text{T}$, $\forall m \in \mathcal{M}$, where the APV optimization is performed by Algorithm~\ref{al_1}.}
		
	\item {6DMA-orientation: This scheme implements the 6DMA with orientation optimization only at each AP with fixed position, i.e., $\mathbf q_m=[0,0,0]^\text{T}$, $\forall m \in \mathcal{M}$, where the AOM optimization is performed by Algorithm~\ref{al_2}.}
	
	\item {FA: Each AP is equipped with a single FA with its position and orientation set equal to that of 6DMA-orientation/6DMA-position, respectively,  and the receive combining matrix is designed based on the MMSE receiver.}
\end{itemize}

\begin{figure}[t]
	\begin{center}
		\includegraphics[width=\figwidth cm]{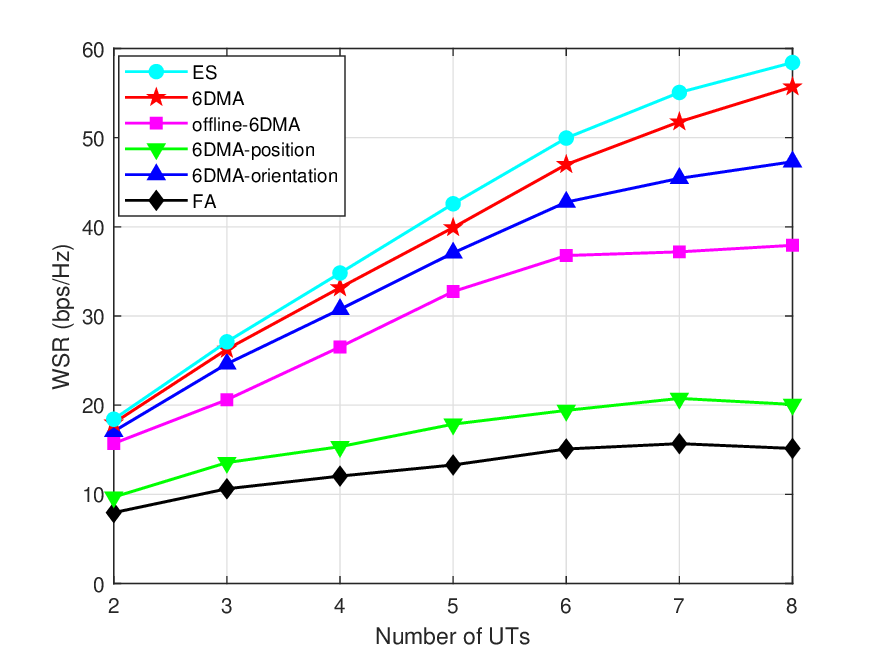}
		\caption{Comparison of the WSR between the proposed and benchmark schemes versus number of UTs.}
		\label{fig:RvsK}
	\end{center}
	\vspace{-0.2 in}
\end{figure}

\begin{figure}[t]
	\begin{center}
		\includegraphics[width=\figwidth cm]{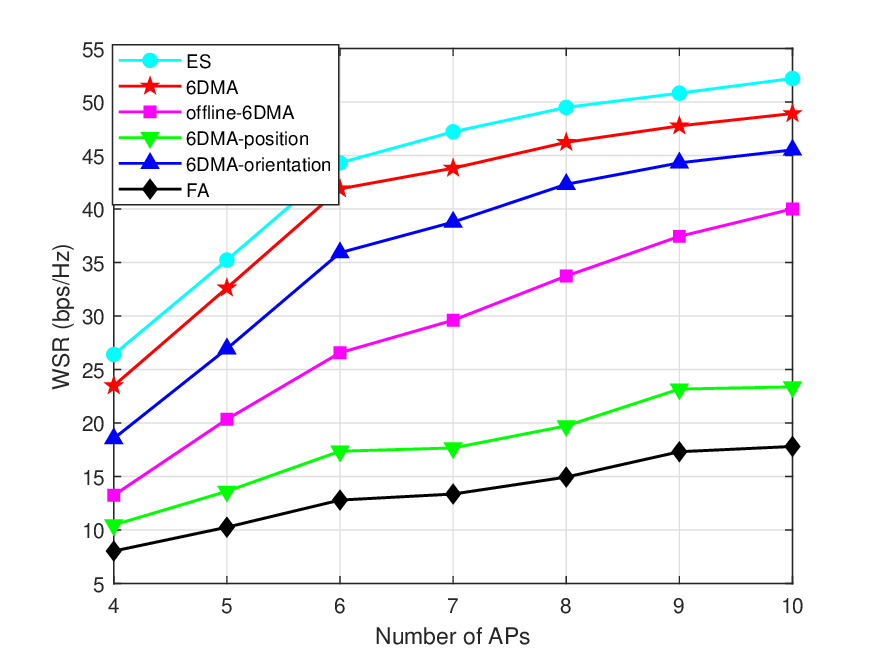}
		\caption{Comparison of the WSR between the proposed and benchmark schemes versus number of APs.}
		\label{fig:RvsM}
	\end{center}
\end{figure}

\begin{figure}[t]
	\begin{center}
		\includegraphics[width=\figwidth cm]{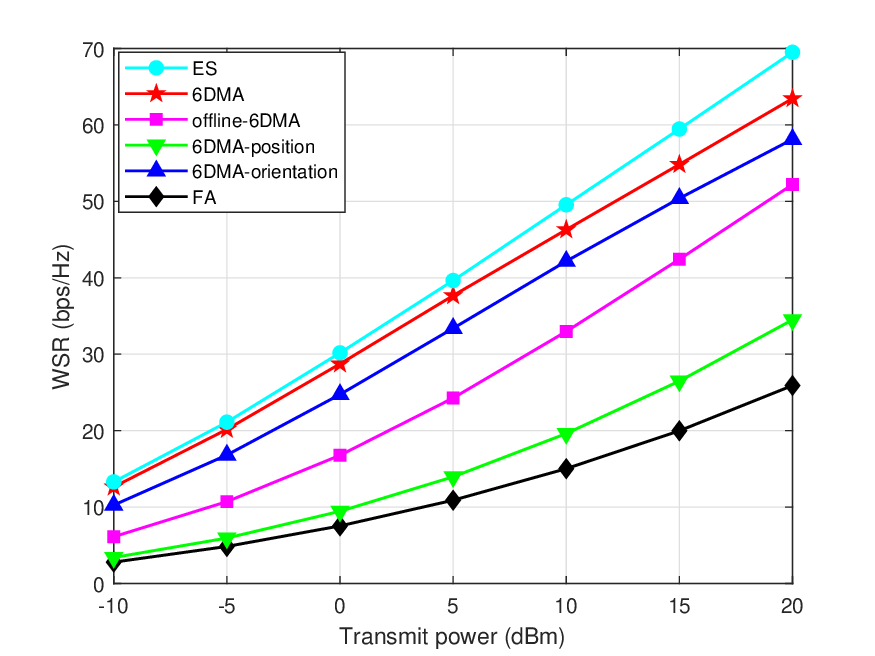}
		\caption{Comparison of the WSR between the proposed and benchmark schemes versus transmit power.}
		\label{fig:RvsP}
	\end{center}
	\vspace{-0.2 in}
\end{figure}

\begin{figure}[t]
	\begin{center}
		\includegraphics[width=\figwidth cm]{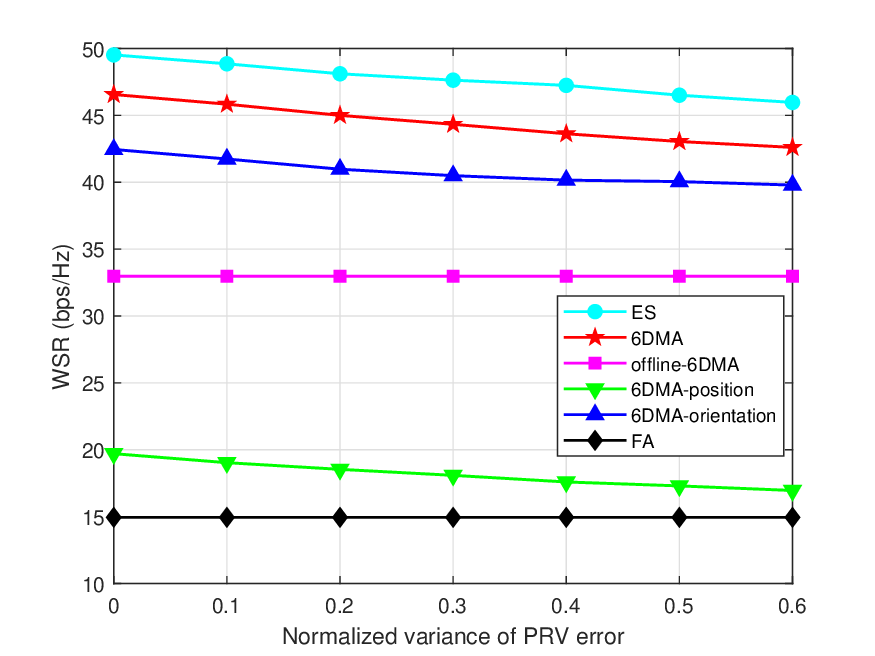}
		\caption{Comparison of the WSR between the proposed and benchmark schemes versus normalized variance of PRV error.}
		\label{fig:RvsPRV}
	\end{center}
\end{figure}

\begin{figure}[t]
	\begin{center}
		\includegraphics[width=\figwidth cm]{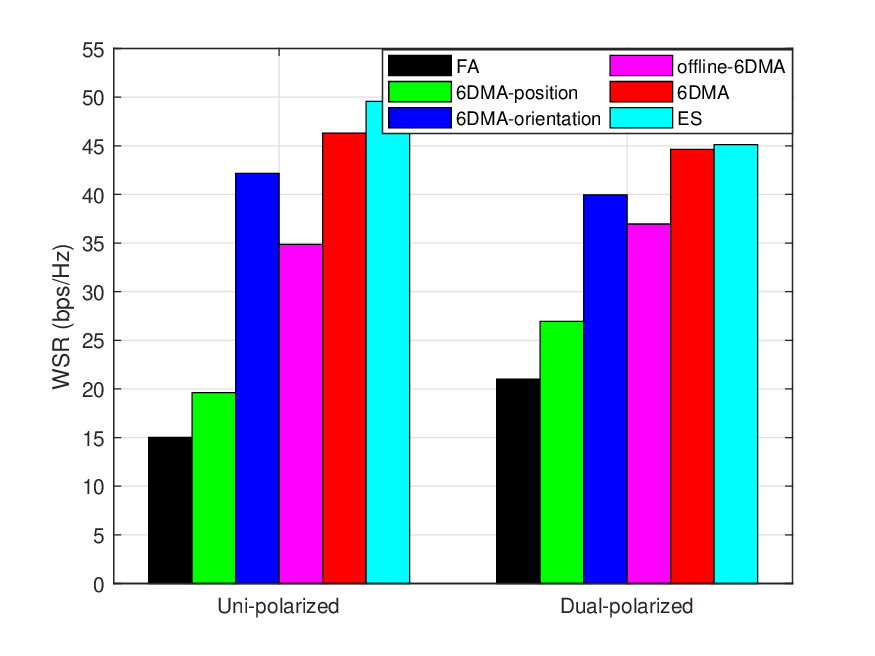}
		\caption{Comparison of the WSR between the proposed and benchmark schemes under the uni-polarized and dual-polarized cases.}
		\label{fig:dual}
	\end{center}
\end{figure}

In Fig.~\ref{fig:RvsK}, we compare the WSR for different schemes versus the number of UTs. It can be seen that the proposed schemes outperform all other benchmark schemes and approach the performance of the ES scheme. With the increasing number of UTs, the WSR increases with a decreasing speed since more UTs will lead to more severe multiuser interference. However, thanks to the ability of adjusting the antennas' positions/orientations to the regions/directions with lower channel correlation among UTs, 6DMA can efficiently suppress the multiuser interference and attain more desired channel power gain compared to the FA.
Notably, the offline-6DMA scheme is significantly superior to the FA scheme, which validates that the proposed offline solution for APV and AOM can still achieve considerable performance improvement without the need of frequent antenna movement.

Fig.~\ref{fig:RvsM} compares the WSR for different schemes versus the number of APs. As can be observed, the proposed 6DMA schemes are still superior to 
all other benchmark schemes except for the ES scheme. In addition, the WSRs  substantially increase with the number of APs for all schemes. To be specific, by increasing the number of APs by twofold, the rate performance improvement is nearly proportional. Additionally, the proposed 6DMA scheme can achieve comparable performance to the ES scheme, while requiring much less computational complexity and overhead. 
Note that with the conventional fixed-position and fixed-orientation antennas, the FA scheme performs the worst for all settings since it fails to exploit the wireless channels' spatial DoFs to increase the channel power gain or decrease the multiuser interference.

Fig.~\ref{fig:RvsP} compares the WSR for different schemes versus the transmit power. It can be observed that the proposed scheme consistently surpassess
all other benchmark schemes except for the ES scheme. This indicates that the optimization of APV and AOM at APs can save the transmit power of each UT under the same WSR requirements. Moreover, the performance gaps between the proposed schemes and the FA scheme considerably increase with the transmit power. This is because in the high signal-to-noise ratio (SNR) regime, the multiuser communication performance is limited by interference, where the decrease in channel correlation for multiple users achieved by antenna movement optimization can significantly reduce their interference and thus improves the spatial multiplexing performance more pronouncedly. Moreover, although the offline-6DMA scheme sacrifices some rate performance compared to its online counterpart, the antenna movement overhead is significantly reduced and its interference mitigation and channel power enhancement are more effective compared to the FA scheme. This result verifies that the offline-6DMA scheme provides a practical solution for implementing the 6DMA-enhanced multi-AP coordination communication systems with low control overhead and energy consumption.


To analyze the impact of imperfect CSI on the optimization of APV and AOM, we compare the WSR in Fig.~\ref{fig:RvsPRV} for different schemes versus the PRV error. It should be emphasized that the receive combining matrices for both 6DMA and FA systems are still obtained based on perfect CSI  for fairness. The PRV errors are defined as the differences between actual PRVs and their estimated counterparts, which are assumed to be i.i.d. CSCG randomly distributed, i.e., $a_{k,m}^l - \hat{a}_{k,m}^l \sim \mathcal{CN}(0,\xi)$ with  $\hat{a}_{k,m}^l$ and $\xi$ denoting estimated PRVs and the variance of the normalized PRV error, respectively. From Fig.~\ref{fig:RvsPRV}, we can see that although the optimizations of APV and AOM both suffer from performance degradation with the increase of PRV error, the proposed 6DMA schemes can still achieve higher performance gain compared to the FA scheme. In addition, the performance gap between the 6DMA and offline-6DMA schemes narrows, which demonstrates the offline-designed APV and AOM based on statistical CSI can attain a robust performance improvement against the imperfect instantaneous CSI.

Fig.~\ref{fig:dual} compares the WSR for different schemes under the uni-polarized and dual-polarized cases. To provide a fair comparison, the number of APs of the uni-polarized schemes is double to that of the dual-polarized schemes. As observed, for both in uni-polarized and dual-polarized
cases, the optimization of APV and AOM can achieve a considerable rate performance improvement, which validates the superiority of 6DMAs. In addition, the uni-polarized scheme is superior to the dual-polarized scheme since the former has higher spatial DoFs in independent movement for each antenna. However, the dual-polarized scheme only needs to control half the number of APs and achieves a performance comparable to the uni-polarized scheme, which effectively reduces the overhead of AP deployment and antenna movement.
\section{Conclusion}\label{sec_Conclusion}
In this paper, we proposed a 6DMA-enhanced multi-AP coordination system to improve the multiuser uplink communication performance. We first characterized the collective channels between distributed APs and multiple UTs, considering the impacts of antennas' positions and orientations on path responses and antenna gains, respectively. Then, the APV and AOM at APs as well as their receive combining matrix were jointly optimized to maximize the WSR of UTs, under the constraints on antenna movement regions. To address this non-convex optimization problem efficiently, we transformed it into a more tractable Lagrangian dual problem by introducing auxiliary variables. Subsequently, an AO-based algorithm was developed
by iteratively optimizing the APV and AOM, where the optimal receive combining was obtained by the MMSE receiver and the suboptimal APV and AOM were obtained by applying the SCA technique and Riemannian manifold optimization-based algorithm, respectively. In addition, we further extended the proposed system from uni-polarized to dual-polarized mode for AP's antennas and proposed an offline solution for APV and AOM design based on statistical CSI. 
Simulation results validated that compared to conventional FA-based systems, the proposed 6DMA-based system can significantly improve the rate performance by exploiting the spatial DoFs in antenna position and orientation optimization with both uni-polarized and dual-polarized antennas. In addition, with significantly reduced overhead of antenna  movement, the offline-6DMA can still attain a considerable performance gain over the FA schemes and thus serves as an efficient solution for implementing 6DMA-enhanced multi-AP coordination systems in practice.

\appendices
\section{Derivations of $\nabla\bar F(\mathbf{q}_m)$ and $\nabla^2\bar F(\mathbf{q}_m)$}\label{app:A}
\allowdisplaybreaks[4]
For ease of exposition, we define
\begin{equation}\label{b_mk}
	\small	
\mathbf b_{k,m} \triangleq(\mathbf{\Lambda}_{k,m}-\mathbf{C}_{k,m})\mathbf{f}_{k,m}(\mathbf q_m^i)+{c}_{k,m}{{\bf{G}}_{k,m}}({{\bf{A}}_m}){{\bf{a}}_{k,m}}.\notag
\end{equation}
Then, by denoting the $l$-th entry of $\mathbf b_{k,m}$ as $b_{k,m}^l$, $F(\mathbf{q}_m)$ can be written as	
\begin{small}
\begin{align*}
\small
\bar F(\mathbf{q}_m) &=\sum\limits_{k \in {\mathcal K}}
2\Re\left\{\mathbf{f}_{k,m}^{\text H}(\mathbf q_m)\mathbf b_{k,m}\right\}\\
&=\sum\limits_{k \in {\mathcal K}}2\Re\left\{\sum\limits_{l=1}^{L_{k,m}}|b_{k,m}^l|e^{j(-\frac{2\pi}{\lambda}\mathbf d(\psi_{k,m}^l)^{\text T}\mathbf q_m+\angle{b_{k,m}^l})}\right\}\\
&=\sum\limits_{k \in {\mathcal K}}\sum\limits_{l=1}^{L_{k,m}}2|b_{k,m}^l|\cos(\Upsilon_{k,m}^l(\mathbf q_m)),\notag	
\end{align*}
\end{small}
with $\Upsilon_{k,m}^l(\mathbf q_m) \triangleq 2\pi\mathbf d(\psi_{k,m}^l)^{\text T}\mathbf q_m/\lambda-\angle{b_{k,m}^l}$. The gradient vector and Hessian matrix of $\bar F(\mathbf{q}_m)$ are given by $\nabla \bar F(\mathbf{q}_m)=\left[\frac{\partial \bar F(\mathbf{q}_m)}{\partial x_m},\frac{\partial \bar F(\mathbf{q}_m)}{\partial y_m},\frac{\partial \bar F(\mathbf{q}_m)}{\partial z_m}\right]^{\text T}$
with 
\begin{small}
\begin{align}\label{delta_F}
	\frac{\partial \bar F(\mathbf{q}_m)}{\partial x_m}
	&= -\frac{4\pi}{\lambda}\sum\limits_{k \in {\mathcal K}}\sum\limits_{l=1}^{L_{k,m}}|b_{k,m}^l|[\mathbf d(\psi_{k,m}^l)]_1\sin(\Upsilon_{k,m}^l(\mathbf q_m)),\notag\\
	\frac{\partial \bar F(\mathbf{q}_m)}{\partial y_m}
	&= -\frac{4\pi}{\lambda}\sum\limits_{k \in {\mathcal K}}\sum\limits_{l=1}^{L_{k,m}}|b_{k,m}^l|[\mathbf d(\psi_{k,m}^l)]_2\sin(\Upsilon_{k,m}^l(\mathbf q_m)),\notag\\
	\frac{\partial \bar F(\mathbf{q}_m)}{\partial z_m}
	&= -\frac{4\pi}{\lambda}\sum\limits_{k \in {\mathcal K}}\sum\limits_{l=1}^{L_{k,m}}|b_{k,m}^l|[\mathbf d(\psi_{k,m}^l)]_3\sin(\Upsilon_{k,m}^l(\mathbf q_m)),
\end{align}
\end{small}
and 
\begin{equation}
	\small
	\nabla^2 \bar F(\mathbf{q}_m)=	
\begin{bmatrix}
	\frac{\partial^2 \bar F(\mathbf{q}_m)}{\partial x_m^2}
	&\frac{\partial^2 \bar F(\mathbf{q}_m)}{\partial x_m \partial y_m}  
	& \frac{\partial^2 \bar F(\mathbf{q}_m)}{\partial x_m \partial z_m}\\
	\frac{\partial^2\bar F(\mathbf{q}_m)}{\partial y_m \partial x_m }
	&\frac{\partial^2 \bar F(\mathbf{q}_m)}{\partial y_m^2}  
	&\frac{\partial^2\bar F(\mathbf{q}_m)}{\partial y_m \partial z_m } \\
	\frac{\partial^2\bar F(\mathbf{q}_m)}{\partial z_m \partial x_m }
	&\frac{\partial^2\bar F(\mathbf{q}_m)}{\partial z_m \partial y_m }  &\frac{\partial^2\bar F(\mathbf{q}_m)}{\partial z_m^2}
\end{bmatrix},
\end{equation}
with 
\begin{small}
\begin{align*}
		&\frac{\partial^2 \bar F(\mathbf{q}_m)}{\partial x_m^2}
		= -\frac{8\pi^2}{\lambda^2}\sum\limits_{k \in {\mathcal K}}\sum\limits_{l=1}^{L_{k,m}}|b_{k,m}^l|[\mathbf d(\psi_{k,m}^l)]_1^2\cos(\Upsilon_{k,m}^l(\mathbf q_m)),\\	
		&\frac{\partial^2 \bar F(\mathbf{q}_m)}{\partial y_m^2}
		= -\frac{8\pi^2}{\lambda^2}\sum\limits_{k \in {\mathcal K}}\sum\limits_{l=1}^{L_{k,m}}|b_{k,m}^l|[\mathbf d(\psi_{k,m}^l)]_2^2\cos(\Upsilon_{k,m}^l(\mathbf q_m)),\\	
		&\frac{\partial^2 \bar F(\mathbf{q}_m)}{\partial z_m^2}
		= -\frac{8\pi^2}{\lambda^2}\sum\limits_{k \in {\mathcal K}}\sum\limits_{l=1}^{L_{k,m}}|b_{k,m}^l|[\mathbf d(\psi_{k,m}^l)]_3^2\cos(\Upsilon_{k,m}^l(\mathbf q_m)),\\		
		&\frac{\partial^2 \bar F(\mathbf{q}_m)}{\partial x_m \partial y_m}
		=\frac{\partial^2 \bar F(\mathbf{q}_m)}{ \partial y_m\partial x_m}= \\
		&-\frac{8\pi^2}{\lambda^2}\sum\limits_{k \in {\mathcal K}}\sum\limits_{l=1}^{L_{k,m}}|b_{k,m}^l|[\mathbf d(\psi_{k,m}^l)]_1[\mathbf d(\psi_{k,m}^l)]_2\cos(\Upsilon_{k,m}^l(\mathbf q_m)), \notag \\ 	
		&\frac{\partial^2 \bar F(\mathbf{q}_m)}{\partial x_m \partial z_m}
		=\frac{\partial^2 \bar F(\mathbf{q}_m)}{ \partial z_m\partial x_m}= \\
		&-\frac{8\pi^2}{\lambda^2}\sum\limits_{k \in {\mathcal K}}\sum\limits_{l=1}^{L_{k,m}}|b_{k,m}^l|[\mathbf d(\psi_{k,m}^l)]_1[\mathbf d(\psi_{k,m}^l)]_3\cos(\Upsilon_{k,m}^l(\mathbf q_m)), \notag \\
		&\frac{\partial^2 \bar F(\mathbf{q}_m)}{\partial y_m \partial z_m}
		=\frac{\partial^2 \bar F(\mathbf{q}_m)}{ \partial z_m\partial y_m}= \\
		&-\frac{8\pi^2}{\lambda^2}\sum\limits_{k \in {\mathcal K}}\sum\limits_{l=1}^{L_{k,m}}|b_{k,m}^l|[\mathbf d(\psi_{k,m}^l)]_2[\mathbf d(\psi_{k,m}^l)]_3\cos(\Upsilon_{k,m}^l(\mathbf q_m)). \\    
\end{align*}
\end{small}
Since we have
\begin{equation}
\small
||\nabla^2 \bar F(\mathbf{q}_m)||_2^2\leq ||\nabla^2 \bar F(\mathbf{q}_m)||_F^2\leq 9\left(\frac{8\pi^2}{\lambda^2}\sum\limits_{k \in {\mathcal K}}\sum\limits_{l=1}^{L_{k,m}}|b_{k,m}^l|\right)^2\notag
\end{equation}
and $||\nabla^2 \bar F(\mathbf{q}_m)||_2\mathbf I_3\succeq \nabla^2 \bar F(\mathbf{q}_m)$, $\delta_{m}$ is thus selected as
\begin{equation}\label{delta_m}
	\small
\delta_{m}=\frac{24\pi^2}{\lambda^2}\sum\limits_{k \in {\mathcal K}}\sum\limits_{l=1}^{L_{k,m}}|b_{k,m}^l|,	
\end{equation}
which guarantees $\delta_{m}\mathbf I_3\succeq ||\nabla^2 \bar F(\mathbf{q}_m)||_2\mathbf I_3\succeq \nabla^2 \bar F(\mathbf{q}_m)$.

\bibliographystyle{IEEEtran}
\bibliography{IEEEabrv,MA}
\end{document}